\documentclass[12pt]{article}
\usepackage{amsmath}
\usepackage{graphicx,psfrag,epsf}
\usepackage{enumerate}
\usepackage{url} % not crucial - just used below for the URL
\usepackage{float}
\usepackage[table,xcdraw,dvipsnames]{xcolor}

\usepackage{cite}
\usepackage{listings}
\usepackage{multirow}
\usepackage{lipsum}
\usepackage{amsfonts}
\usepackage{cite}      % Written by Donald Arseneau

\usepackage{graphicx}   % Written by David Carlisle and Sebastian Rahtz

\usepackage{amsthm}
\RequirePackage[colorlinks,citecolor=blue,urlcolor=blue]{hyperref}
\usepackage{amssymb,calc}
\usepackage{thmtools}
\usepackage{mathrsfs}
\usepackage{mathtools}
\usepackage{bm}
\usepackage{bbm}

\usepackage{enumerate}
\usepackage{rotating}
\RequirePackage{cleveref}
\usepackage{hyphenat}
 \usepackage[numbers]{natbib}
\usepackage{caption,subcaption}
\usepackage{tikz}
\usepackage{pgfplots}

\usepackage[T1]{fontenc}
\usepackage{textcomp} % for \textquotesingle
\usepackage{lmodern}

\usepackage[noend]{algpseudocode}
\usepackage{algorithm}
\makeatletter
\renewcommand{\ALG@name}{Algorithm}
\makeatother

\usepackage[framemethod=TikZ]{mdframed}

\newcommand{\prompt}[1]{
\begin{mdframed}[topline=false, bottomline=false, leftline=true, rightline=false, innertopmargin=0pt, innerbottommargin=0pt, innerrightmargin=0pt, innerleftmargin=10pt, leftmargin=3em, rightmargin=3em, linewidth=2pt,linecolor=black, skipabove=12pt, skipbelow=7pt]
\begin{flushleft}
#1
\end{flushleft}
\end{mdframed}
}

\pgfplotsset{width=10cm}

\tikzset{declare function={gamma(\x)=sqrt(2*pi)*\x^(\x-0.5)*exp(-\x)*exp(1/(12*\x));}}
\tikzset{declare function={tpdf(\x,\nu)=gamma(0.5*(\nu+1))/(sqrt(pi*\nu)*gamma(\nu/2))*(1+\x^2/\nu)^(-(\nu+1)/2);}}
\tikzset{declare function={invgampdf(\x,\a,\b)=(\b/\x)^\a/\x/gamma(\a)*exp(-\b/\x);}}

\newcommand{\nhphantom}[1]{\ifmmode\settowidth{\dimen0}{$#1$}\else\settowidth{\dimen0}{#1}\fi\hspace*{-\dimen0}}

\usetikzlibrary{patterns}
\tikzset{
	hatch distance/.store in=\hatchdistance,
	hatch distance=5pt,
	hatch thickness/.store in=\hatchthickness,
	hatch thickness=0.5pt,
}

\newcommand{\wh}[1]{\widehat{#1}}

\definecolor{pink}{rgb}{0.9, 0.17, 0.31}

\def\C {\,|\:}

\renewcommand\d{\mathrm d}

\newcommand\R{\mathbb R}
\renewcommand\b{\bm{\beta}}

\newcommand{\wt}[1]{\widetilde{#1}}

\newtheorem{theorem}{Theorem}

\renewcommand{\nhphantom}[1]{\ifmmode\settowidth{\dimen0}{$#1$}\else\settowidth{\dimen0}{#1}\fi\hspace*{-\dimen0}}

\numberwithin{equation}{section}

\crefname{thm}{Theorem}{Theorems}
\crefname{prop}{Proposition}{Propositions}
\crefname{lem}{Lemma}{Lemmas}
\crefname{coro}{Corollary}{Corollaries}
\crefname{add}{Addendum}{Addendums}
\crefname{asm}{Assumption}{Assumptions}
\crefname{alg}{Algorithm}{Algorithms}
\crefname{proc}{Procedure}{Procedures}
\crefname{exe}{Exercise}{Exercises}
\crefname{exa}{Example}{Examples}
\crefname{prob}{Problem}{Problems}
\crefname{section}{Section}{Sections}
\crefname{subsection}{Section}{Sections}
\crefname{appendix}{Appendix}{Appendices}

\def\argmin{\mathop{\arg\min}}

\allowdisplaybreaks[3]

\usepackage{algorithm}
\usepackage{algpseudocode}
\usepackage{amsmath}

\begin{document}

\def\spacingset#1{\renewcommand{\baselinestretch}%
{#1}\small\normalsize} \spacingset{1}

%%%%%%%%%%%%%%%%%%%%%%%%%%%%%%%%%%%%%%%%%%%%%%%%%%%%%%%%%%%%%%%%%%%%%%%%%%%%%%

%\if1\blind
%{
	\title{\sf  AI-Powered Bayesian Inference}
	\author{ Sean O'Hagan\footnote{Sean  O'Hagan is a 4th year PhD student at the Department of Statistics of the University of Chicago.} \,\, and \,\,  Veronika Ro\v{c}kov\'{a}\footnote{Veronika Ro\v{c}kov\'{a} is the Bruce Lindsay Professor of Econometrics and Statistics in the Wallman Society of Fellows at the Booth School of Business at the University of Chicago. 
	 The author would like to thank Tijana Zrnic for pointing out a reference to catalytic priors which was a catalyst for this research. 
	This research was supported by the NSF (DMS: 1944740).}
			}
	\maketitle
%} \fi

\bigskip
\begin{abstract}
The advent of Generative Artificial Intelligence (GAI) has heralded an inflection point that has changed how society thinks about knowledge acquisition.
While GAI cannot be fully trusted for decision-making, it may still provide valuable information that can be integrated into a decision pipeline.
Rather than seeing the lack of certitude and  inherent randomness of GAI as a problem, we view it as an opportunity. Indeed, variable answers to given prompts can be leveraged to  construct a prior distribution which reflects assuredness of AI predictions. This prior distribution may be combined with tailored datasets for a fully Bayesian analysis with an AI-driven prior.
In this paper, we explore such a possibility within a non-parametric Bayesian framework. The basic idea  consists of assigning a Dirichlet process prior distribution on the data-generating distribution with AI generative model as its baseline.  Hyper-parameters of the prior can be tuned out-of-sample to assess the informativeness of the AI prior.
Posterior simulation is achieved by computing a suitably randomized functional on an augmented data that consists of observed (labeled) data as well as fake data whose labels have been imputed using AI. This strategy can be parallelized and rapidly produces iid samples from the posterior by optimization as opposed to sampling from conditionals. Our method enables (predictive) inference and uncertainty quantification leveraging AI predictions in a coherent probabilistic manner.
\end{abstract}

\noindent%
\vfill

\newpage
\spacingset{1.45} % DON'T change the spacing!

\section{Introduction}
Due to its ability to synthesize information from various sources,  Generative Artificial Intelligence (GAI) is quickly becoming a   source of  knowledge for many of its users.
However,  the practical utility of GAI models largely depends on the user's understanding of their mechanistic (probabilistic) properties.  Generative models simply
produce stochastic responses to prompts, with the level of randomness influenced by both the prompt's specificity and the AI's ability and confidence in providing an accurate answer. 
 This inherent randomness raises questions about the extent to which important
decisions can be based  on a single AI output answer. We argue that  the variability in the answers themselves  offers an opportunity to generate (a) a random prior guess at the correct answer, and (b) probabilistic predictions via AI-induced distributions obtained by repeated prompting.
While our society has  resisted surrendering important decision-making to artificial intelligence,  AI predictive systems may serve as a useful primer for further analysis. 
This work explores the possibility of  
 %AI predictions   may result in misleading recommendations not only due to the  inherent randomness but also possible 
%While delegating decisions entirely to  AI intelligence has been approached with reluctance,
  articulating prior information through  AI data augmentation  for a fully Bayesian analysis.

Our setup consists of independent labeled data $\mathcal D_n= \{(Y_i,\bm X_i)\}_{i=1}^n$  which are tailored to a specific question regarding a parameter of interest $\theta_0$. For example, we will later analyze a dermatology dataset where the label is one of six distinct but closely overlapping skin conditions within the group of Erythemato-Squamous Diseases (ESDs) \citep{wang2025identifying}. While the parameter of interest $\theta_0$  may index a statistical model (including deep learning models involving high-dimensional $\theta_0$), we regard it more generally as a minimizer of a certain loss function \citep{bissiri2016general}.  Our goal is to understand  how generative AI can be used for prior elicitation to conduct  fully Bayesian inference on  $\theta_0$ as well as predictive inference on   $Y_i^*$ given $\bm X^*_i\notin\mathcal D_n$.

This work is based on the premise that generative models produce synthetic  data which can be converted into (informative) priors.  The idea of using imaginary training data for prior construction is nearly as old as Bayesian statistics itself, dating to at least Laplace in the 18th century \citep{good}.
In the context of   Bayes factor model comparisons, intrinsic priors \citep{berger1996intrinsic} result from converting an improper uninformative prior into a proper posterior on a sample of  a ``minimal" training size.
Arithmetic and geometric mean aggregates of Bayes factors under all plausible training data subsets approximately correspond to a Bayes factor under the so-called ``intrinsic prior".
The expected-posterior prior \citep{perez2002expected} is a special case  which results from averaging  posterior distributions, given imaginary data, over all imaginary data arising from a suitable predictive measure (for example predictive distribution under a simpler model or empirical distribution of the observed data).
Besides intrinsic and expected-posterior priors, data augmentation ideas for prior constructions for model determination have been explored by many others, including \cite{spiegelhalter1982bayes,gelfand1992model,laud1996predictive} or \citep{ohagan1995fractional}.
We are not necessarily concerned with objective prior  elicitation for  model selection but rather with subjective prior articulation for actual inference about $\theta_0$ under one posited model.

Only very few priors have had as much practical and theoretical impact as the Zellner's $g$-prior \citep{zellner1986assessing}.
Motivated by imaginary data, this prior adopts the covariance structure from  observed data to facilitate conjugate analysis in normal regression.  Inspired by \citep{diaconis1979conjugate},  \citep{ibrahim2000power,chen2003conjugate} propose  predictive elicitation of a proper conjugate prior for generalized linear models based on observable quantities (historical data) which serve as a prior guess at  observable outcomes. Similar data augmentation (DA) priors, that have  the same form as the likelihood, have also been studied by \citep{bedrick1996new} who considered a broader class of conditional mean priors  arguing that it is easier to elicit prior information on means of observables rather than on regression coefficients. Related ideas occurred in specialized contexts including proportional hazards regression \citep{greenland2001data}.  A more recent  incarnation of DA priors  is the catalytic prior  \citep{huang2020catalytic} that involves simulated data from a posterior predictive distribution under a simpler (donor) model trained on $\mathcal D_n$. The catalytic prior is then constructed from the, suitably down-weighted,  fake training samples so that it is conjugate with the posited model. While the catalytic prior appears conceptually related to the expected-posterior prior \citep{perez2002expected,fouskakis2018power,fouskakis2016power,neal2001transferring} as well as power-priors \citep{chen2003conjugate}, it originates from a different set of desiderata.
With complex models and small datasets,    likelihood-based analysis can be unstable or infeasible and may  be enriched by
augmenting the observed data with imaginary data generated from a posterior predictive under a simpler model \citep{huang2020catalytic,neal2001transferring}. This results in a posterior distribution that is pulled  towards the posterior  under the simpler model, resulting in estimates and predictions with potentially better properties.  Our work has been motivated by  catalytic priors\footnote{This reference was pointed out to one of the authors by Tijana Zrnic during her visit at Booth.} in the sense that we also view  generative AI   as a posterior predictive simulator  but we approach AI prior articulation differently.
First,  we use simulations from  generative AI models trained on massive datasets that are different from the proprietary observed data $\mathcal D_n$. Catalytic priors use  simulations from posterior predictive distributions under models trained on  $\mathcal D_n$, inherently using $\mathcal D_n$ twice. Second, we consider a non-parametric Bayesian inference framework by constructing a prior  for $F_0$, the distribution function underlying independent realizations inside $\mathcal D_n$, as opposed to  a prior on $\theta_0$. Shifting focus from inference on an unknown parameter  $\theta_0$ to inference on an unknown data-generating distribution  $F_0$ allows generative AI output to be harnessed in a more transparent way.  Indeed,  generative AI  can be leveraged to produce imaginary data   $\mathcal D^*_m=\{(Y_i^*,\bm X_i^*)\}_{i=1}^m$ which could be regarded as samples from the prior on $F_0$. We align with  \citep{chen2003conjugate} who state that ``{\em it is easier to think of observable quantities when eliciting priors, rather than
specifying priors for regression parameters directly, since parameters are always unobserved}."  However, unlike \citep{chen2003conjugate}, we focus directly on non-parametric inference on $F_0$  which  obviates the need to commit to a particular model and  breaks the precarious codependence between the (conjugate) prior and the model.

%Our construction is predictive in nature and focuses on observable quantities, it is based on specifying a prior prediction y0 for the response
%vector, and a scalar precision parameter a0 which quantifies one?s prior belief in y0.
Our idea is extremely simple. We view generative AI as a base distribution for a Dirichlet process prior on $F_0$ \citep{ferguson1973bayesian}.  
Inference on $\theta_0$ can be accomplished through the loss-based posterior framework of \citep{bissiri2016general,lyddon2019general, chamberlain1996nonparametric}
  where the parameter of interest $\theta_0$ is a functional of $F_0$. The Dirichlet process prior admits an embarrassingly parallel posterior bootstrap algorithm \citep{lyddon2019general,fong2019scalable} that generates independent and exact samples from the nonparametric posterior distribution over the data-generating process.
Since the AI prior on $\theta_0$ is  defined only implicitly through an AI prior on $F_0$, Markov chain Monte Carlo (MCMC) analysis that relies on sampling from conditionals is not immediately unavailable. Instead of obtaining dependent samples using MCMC, we obtain independent posterior samples by optimization of  suitably randomized objective functions along the lines of \cite{fong2019scalable}. This strategy allows us to obtain posterior distributions of functionals as well as posterior predictive distributions. These distributions can be used to report (predictive) uncertainty  in optimization-based machine learning techniques (such as random forests or deep learning) for which uncertainty quantification has been challenging or unavailable. Our work has been inspired by the prediction-powered inference framework \citep{angelopoulos2023prediction,ji2025predictions,zrnic2024cross} for constructing valid confidence intervals and  $p$-values that allows for AI systems to be used for data augmentation. {The prediction-powered inference (PPI) framework posits estimators (or confidence sets) for minimizers of convex objectives that leverage predictions from a black-box AI model on additional unlabeled data in order to aid estimation. PPI does this by making use of a rectifier which relates AI prediction error to bias and employing a strategy reminiscent of de-biasing.} One aspect of our work can be viewed as a Bayesian counterpart to this framework.

The paper is structured as follows. In Section \ref{sec:Bayesian_AI} we review  various   prior elicitation methods using AI data augmentation in likelihood-driven Bayesian inference.  Section \ref{sec:likelihood_free} presents AI prior elicitation in loss-function driven Bayesian inference. Section \ref{sec:theory} presents theory and hyperparameter calibration strategies. Section \ref{sec:Applications} shows real data applications including  AI-assisted disease diagnosis, astrophysics, and genetics. Section \ref{sec:Discussion} wraps up with a discussion.

\section{AI-Powered Bayesian Inference}\label{sec:Bayesian_AI}
Suppose that we observe labeled data $\mathcal D_n =\{(Y_i,\bm X_i)\}_{i=1}^n$ and want to perform a supervised analysis involving a parameter of interest $\theta_0\in \Theta\subseteq\R^d$ as well as predictive inference about $Y_{new}$ given $\bm X_{new}=\bm x_{new}$ where $(Y_{new},\bm X_{new})\notin \mathcal D_n$. In addition to observations $\mathcal D_{n}$, we have access to a (stochastic) black-box predictive model $\hat{\mu}_{AI}(\bm x)$  which  generates random labels $\wt Y_i=\hat{\mu}_{AI}(\bm x)$ when prompted by $\bm x$.
One can regard $\hat{\mu}_{AI}(\bm x)$ as an implicit simulator from a predictive distribution of a complex model (e.g. a large language model underlying generative AI) trained on massive data $\mathcal{D}_{AI}$ that is unavailable to the user and different from $\mathcal D_n$.

To conduct  predictive inference, a Bayesian forecaster would typically issue a posterior predictive distribution
\begin{equation}
\pi(Y_{new}\C \bm x_{new}, \mathcal D_n)=\int \pi(Y_{new}\C \bm x_{new},\theta)\pi(\theta\C \mathcal D_n)\d\theta\label{eq:predictive_inference}
\end{equation}
 based on the posterior distribution of $\theta$ given $\mathcal D_n$
$$
\pi(\theta\C \mathcal D_n)\propto \pi(\theta)\pi(\mathcal D_n\C \theta)
$$
under a postulated model $\pi(\mathcal D_n\C \theta)$ and a chosen prior $\pi(\theta)$. 
 There are two conundrums that have given a pause to some practitioners before fully embracing Bayesian inference: (1) the specification of the prior $\pi(\theta)$, and (2) the specification of the model, e.g. $\pi(\mathcal D_{n}\C \theta)=\prod_{i=1}^n\pi(Y_i\C \bm X_i,\theta)$ for independent realizations. While Bayesians have historically faced taunts about (subjective) priors, model choice is  far more consequential in a likelihood-based analysis and should be defended  at least  as fiercely as the prior choice.
 This work  deals with (subjective) prior elicitation based on observable predictions from an AI predictive model.   The issue of model specification will be tackled using a non-parametric framework Section \ref{sec:NP_AI_priors} which allow Statisticians ``{\em to remain honest about their  ability to perfectly model the data}'' \citep{lyddon2019general}.
 
In the following three subsections, we present several approaches for AI prior elicitation that could be used in the context of likelihood-driven Bayesian inference. Our main discussion will center around  nonparametric Bayesian  inference driven by loss function in Section \ref{sec:likelihood_free}.

 \subsection{Likelihood-Driven AI Priors}\label{sec:P}
 While the AI model  is a black box  predictive machine, it implicitly defines a model and a prior if one were willing to assume that $\hat{\mu}_{AI}(\cdot)$ generates  samples from a posterior predictive distribution \eqref{eq:predictive_inference} under  some label distribution $\pi_{AI}(Y\C \bm x,\theta)$, prior $\pi_{AI}(\theta)$  and a training model $\pi(\mathcal{D}_{AI}\C\theta)$.
This argument might be justifiable from the ``prequential" point of view \citep{dawid1992prequential} that focuses solely on predictive distributions (as opposed to models and priors) and argues that the quality of an inference method can truly be gauged  by the quality of its forecasts. We regard AI forecasts as a potentially useful proxy for the true unobserved outcomes.

One hypothetical (but impossible)  strategy of turning  AI knowledge into priors would be to utilize $\pi_{AI}(\theta\C \mathcal D_{AI})\propto \pi_{AI}(\theta)\pi(\mathcal{D}_{AI}\C\theta)$ as a prior $\pi(\theta)$ for   the predictive distribution $\pi(Y_{new}\C \bm x_{new}, \mathcal D_n)$ based on labeled data $\mathcal D_n$. However, we cannot directly access the parameter posterior simulator $\pi_{AI}(\theta\C \mathcal D_{AI})$ from $\hat{\mu}_{AI}(\cdot)$. What we can access, however, is imaginary data $\mathcal D^*_m=\{(Y_i^*,\bm X_i^*)\}_{i=1}^m$ consisting of  predictive imputations from $\hat{\mu}_{AI}(\cdot)$. We explore  data augmentation strategies for parametric priors in Section \ref{sec:Par} and for non-parametric priors in Section \ref{sec:NP_AI_priors}. These approaches
should be distinguished from martingale posteriors \citep{fong2021martingale} which also leverage predictive imputation for posterior computation but in a very different way.
Martingale posteriors are based on large sequences of missing data generated from one-step-ahead  predictive distributions that are continuously updated with newly generated data.
A posterior distribution over a parameter of interest for exchangeable observations is obtained using the Doob's theorem  by computing a functional (such as the mean) of observed data augmented with the  imputed sequence. In contrast, we do not consider predictive imputation  of an (infinite) sequence  of observations  from continuously updated posterior predictive. Rather, we  perform  imputation of a finite number of  fake observations from a  given ``predictive" distribution which does not update with newly generated imaginary data points.

\subsubsection{Power AI Priors}\label{sec:Par}
We align with the insight by  \cite{chen2003conjugate} that ``{\em it is much easier to elicit information about the typical outcome than to attempt the extremely difficult task of eliciting prior knowledge about $\theta$}". Transmitting prior information through data augmentation has a long history in Bayesian statistics \citep{good},  Zellner's $g$-prior being  perhaps the most prominent example.  
If we were to generate fake training data $\mathcal D^*_m=\{(Y_i^*,\bm X_i^*)\}_{i=1}^m$ (using either ${\bm X}_i^*=\bm X_i$ or by sampling with replacement from  $\bm X_i$'s), we can augment  $\mathcal D_n$ with $\mathcal D_m^*$ and apply any (Bayesian) procedure on this joint sample under some posited model $\pi(Y\C\bm X,\theta)$. The contribution of imaginary data could be possibly down-played by raising their contribution to the joint likelihood to a small power $1/\delta>0$. This is the basic premise of data-augmentation priors. The power-prior  \cite{ibrahim2000power, chen2003conjugate} is one of the early examples within the context of generalized linear models, where  $\bm X_j^*=\bm X_j$ for $1\leq j\leq m=n$  and where auxiliary labels $Y_j^*$ serve as a fixed prior guess at $Y_j$ given $\bm X_j$. Following \citep{diaconis1979conjugate},    \cite{chen2003conjugate} construct a conjugate prior  by plugging $\mathcal D_m^*$ into the likelihood of an exponential family model assuming that $Y_j^*$'s are conditionally independent, given a model parameter $\theta$. In addition, each imaginary data point $Y_j^*$ (given $\bm X_j^*=\bm X_j)$ is down-weighted by some small parameter $1/ \delta>0$ and could be interpreted as a prior prediction (or guess) for $E[Y_j\C \bm X_j]$.  Zellner's $g$-prior is a special case in normal regression  where $g=\delta$. The construction in \citep{chen2003conjugate} is related but different from data-augmentation priors of  \citep{bedrick1996new} who allow for $m$ to be different from $n$ and where $\bm X_i^*$ is not necessarily one of the $\bm X_i$'s.  Moreover,  while \cite{chen2003conjugate} assigns the same weight parameter
$\delta$ to $Y_j^*$ that acts as an effective prior sample size for the prior, the framework of 
\citep{bedrick1996new} assigns a weight $w_j^*$ to each new observation $(Y_j^*,\bm X_j^*)$ where $w_j^*$ can be viewed as a possibly fractional number of observations associated with a particular $(Y_j^*,\bm X_j^*)$. We will see later in  Section \ref{sec:NP_AI_priors}  that thinking of $w_j^*$'s as random rather than fixed will correspond to a  Bayesian bootstrap style strategy.   

Power AI priors could be constructed by regarding   AI predictions $Y_j^*=\hat \mu_{AI}(\bm X_j^*)$ as arising from the same model $\pi(Y\C\bm X,\theta)$ as the data $\mathcal D_n$ using either \cite{chen2003conjugate} or  \citep{bedrick1996new} as follows:
$$
\pi_{Power}(\theta)\propto \prod_{i=1}^m \pi(Y_i^*\C\bm X_i^*,\theta)^{1/\delta}\pi_W(\theta)
$$
where $\pi_W(\theta)$ is some baseline working prior. Power priors could be   enhanced  by incorporating the randomness in $\mathcal D_m^*$. Instead of building a prior from one particular realization of $\mathcal D_m^*$, expected-posterior priors \citep{perez2002expected,fouskakis2018power,fouskakis2016power,neal2001transferring}  and catalytic priors \cite{huang2020catalytic} incorporate randomness of $\mathcal D_m^*$ but do so in different ways. We explain the differences below.

\subsubsection{Expected-Posterior AI Priors}\label{sec:EP}

The expected-posterior  AI prior along the lines of Definition 1 in \citep{perez2002expected} could be constructed as a typical power prior after margining out the imaginary data
\begin{equation}
\pi_{EP}(\theta)\propto \int \prod_{i=1}^m \pi(Y_i^*\C\bm X_i^*,\theta)\pi_W(\theta) \pi_{AI}(\mathcal D_m^*)\d \mathcal D_m^*,\label{eq:EP_AI}
\end{equation}
where $\pi_{AI}(\cdot)$ consists of first generating prompts $\bm X^*$ (possibly using observed $\bm X_j$'s)  and labels $Y^*$ from the posterior predictive distribution underlying the simulator $\hat\mu(\bm X^*)$ (as discussed at the beginning of Section \ref{sec:P}). The posterior distribution $\pi(\theta\C \mathcal D_n)$ under the prior \eqref{eq:EP_AI} corresponds to a typical joint posterior under the prior $\pi_{W}(\theta)$ after averaging out   $ \mathcal D_m^*$. Indeed, under the prior \eqref{eq:EP_AI} we have
\begin{equation}
\pi(\theta\C  \mathcal D_n)= \int \pi(\theta\C  \mathcal D_n,\mathcal D_{m}^*)\pi_{AI}(\mathcal D_{m}^*)\d \mathcal D_m^*.\label{eq:post_EP_AI}
\end{equation}
This characterization has a  practical benefit  for posterior simulation from  \eqref{eq:post_EP_AI}.
A Markov chain  $\{ \theta^{(t)}\}_{t=1}$ with a stationary distribution \eqref{eq:post_EP_AI} can be obtained  by generating a joint chain $\{(\theta^{(t)},\mathcal D_{m}^{*(t)}\}_{t=1}^T$ by first refreshing $\mathcal D_m^*$ from $\pi_{AI}(\mathcal D_{m}^*)$ at every MCMC iteration and then, given $\mathcal D_m^*$, generate $\theta^{(t)}$ from the joint posterior. 
Marginally, $\theta^{(t)}$'s would be distributed according to \eqref{eq:post_EP_AI}.
Unlike with power priors, this strategy refreshes the fake data during simulation as opposed to conditioning on them a-priori. A related approach was considered by \cite{neal2001transferring} in the context of exchangeable observations where predictions from a simple donor model, for which prior elicitation was feasible, were transferred to a more complex recipient model through imaginary data. The implications of treating the imaginary data as random as opposed to fixed were explored in \cite{kaji2023metropolis} in the context of contrastive learning for Bayesian computation using the Metropolis-Hastings algorithm.

\subsubsection{Catalytic AI Priors}\label{sec:catalytic}
In catalytic priors \citep{huang2020catalytic},  imaginary data $\mathcal D_m^*$ are generated from a Bayesian predictive distribution under a simple donor model trained on $\mathcal D_n$ for which prior elicitation was easier.  The data $\mathcal D_m^*$ are then plugged into a likelihood representing  a more complex recipient model whose parameters would be difficult to estimate using only $\mathcal D_n$.
 Formally, the catalytic version of an AI prior could be written as
\begin{equation}
\pi_{CAT, m}(\theta)\propto \left(\prod_{i=1}^m \pi(Y_i^*\C\bm X_i^*,\theta)\right)^{\alpha/m}=\exp\left\{\frac{\alpha}{m}\sum_{i=1}^m\log \pi(Y_i^*\C\bm X_i^*,\theta)\right\}\label{eq:prior_CAT}
\end{equation}
for some $\alpha>0$ which regulates the influence of the prior and where $1/m$ performs averaging over the contributions of single imaginary data points $Y_i^*$. A similar idea could be implemented using AI predictions. Unlike catalytic priors, however, generating fresh data $\mathcal D^*_m$ from an AI model precludes from the double use of data $\mathcal D_n$.  The practical implementation of Bayesian analysis with catalytic priors \eqref{eq:prior_CAT} would entail choosing $\alpha$ using some criterion and then simulating very many fake observations $m$ so that the  averaging in \eqref{eq:prior_CAT} performs satisfactory approximation to Monte Carlo integration. Indeed, as $m\rightarrow\infty$ the prior approaches
\begin{equation}
\pi_{CAT, \infty}(\theta)\propto \exp\left\{\alpha\int  \log \pi(Y^*\C\bm X ^*,\theta) \pi_{AI}(Y^*, \bm X^*)\d (Y^*,\bm X^*) \right\}.\label{eq:CAT_inf}
\end{equation}
We highlight an important difference  between \eqref{eq:CAT_inf} and the expected-posterior prior  \eqref{eq:EP_AI}. If we denote the unnormalized expressions in \eqref{eq:CAT_inf} and \eqref{eq:EP_AI} as $f_{CAT,\infty}$ and $f_{EP}$ respectively, then from the Jensen's inequality $E \log X\leq \log E X$, the
population catalytic prior satisfies $f_{CAT,\infty}(\theta)\leq f_{EP}(\theta)$ for all $\theta\in\Theta$, $\alpha\in \mathbb N$ with  $m=\alpha$, $\pi_{W}(\theta)\propto 1$, and $\pi_{AI}(\mathcal D_{m}^*)=\prod_{i=1}^m \pi_{AI}(Y^*_i, \bm X^*_i)$. The expected-posterior prior is more general and allows for information blending from a larger set of imaginary data that are not necessarily iid. In summary,  $m$ in the prior  \eqref{eq:EP_AI} actually corresponds to  the imaginary data sample size represented by $\alpha$ in \eqref{eq:prior_CAT} with $m=\infty$.  %For more general $\alpha>0$, it is a lower bound to a power expected-posterior prior by Fouskakis et al. (2015).

While the expected-posterior prior  \eqref{eq:EP_AI} allows for exact posterior simulation by updating the fake data at each MCMC simulation step by averaging out uncertainty in $\mathcal D_m^*$, catalytic priors \eqref{eq:prior_CAT} compute a different object due to the Jensen's gap. The population version actually corresponds to posteriors  obtained by augmenting a single typical imaginary distribution using estimated moments of $(Y^*,\bm X^*)$. This can be seen, for example, in Gaussian linear regression with unit variance, where  the catalytic prior based on simulations $\mathcal D_{m}^*$ would become
$$
\pi_{CAT,m}=N\left(\hat \theta,\frac{m}{\alpha}(\bm X^{*'}\bm X^*)^{-1}\right)
$$
where $\hat\theta=(\bm X^{*'}\bm X^*)^{-1}\bm X^{*'}\bm Y^* $. If we choose $\bm X^*=\bm X$ where only the labels $Y_i^*$ are subject to predictive imputation, this corresponds to the $g$-prior with  $g=m/\alpha$.
The population catalytic prior would then become $N(\theta^*,\frac{1}{\alpha}\Sigma_{X}^{-1})$, where $\Sigma_X=\lim_{m\rightarrow\infty}\frac{1}{m}\bm X^{*'}\bm X^*$ and $\theta^*=\Sigma_X^{-1}c$, where $c=\lim_{m\rightarrow\infty}\frac{1}{m}\bm X^{*'}\bm Y^*$. While the population catalytic prior plugs moments into the Gaussian prior, the expected-posterior prior is a Gaussian mixture
$$
 \int N\left(\hat \theta,\frac{m}{\tau}(\bm X^{*'}\bm X^*)^{-1}\right)\pi_{AI}(\mathcal D_m^*)\d\mathcal D_{m}^*.
$$

 All of the prior constructions in Section \ref{sec:Par}, \ref{sec:EP} and \ref{sec:catalytic}  force the observed data $\mathcal D_n$ and AI-generated data $\mathcal D_{m}^*$ into a conjugate relationship. This prescription is much stronger than just assuming that the model is well-specified  because it demands that the prior has arrived from the very same model.   We prefer  avoiding the double mis-specification (model and the prior) and will therefore focus on a non-parametric Bayesian analysis based on  AI-informed priors on $F_0$ as opposed to   $\theta_0$.

 \section{ Bayes without  the Likelihood} \label{sec:likelihood_free}

Instead of assuming that there exists $\theta_0$ such that the observed data $\mathcal D_n$ has been independently realized from $ \pi(Y \C \bm X,\theta_0)$, we adopt a non-parametric viewpoint, where the   $\mathcal D_n$  arrives from an iid experiment involving an unknown distribution function $F_0$ for $(Y,\bm X)$. Similarly as in \citep{bissiri2016general, lyddon2019general}, we shift focus from $\theta_0$ to $F_0$. 
The question of prior elicitation will  be tackled by converting observable predictions from an AI model into non-parametric priors on $F_0$.
Suppose that the unknown parameter $\theta_0$ is a solution to the optimization problem
\begin{equation}
\theta_0(F_0)=\argmin_{\theta}\int \ell(\theta,Y,\bm X)\d F_0[(Y,\bm X)],\label{parameter_loss}
\end{equation}
where $\ell(\theta, Y,\bm X)$ is a  loss function  and $F_0$ is the unknown distribution for $(Y,\bm X)$. The parameter of interest is not necessarily tied to a statistical model and is defined more generally  as a minimizer of a population loss under an unknown sampling distribution $F_0$. This parameter may correspond to an actual parameter of a statistical model if one takes $\ell(\theta, Y,\bm X)=-\log\pi(Y\C\bm X, \theta)$.

\subsection{Gibbs AI Priors}
Bissiri et al. \citep{bissiri2016general} formalized a framework for coherent probabilistic updating  of prior beliefs $\pi(\theta)$ about $\theta_0$ from observations $\mathcal D_n$ through a functional 
$
\pi_{GP}(\theta\C \mathcal D_n)\propto \pi(\theta)\exp\left[- w \ell_n(\theta,\mathcal D_n)\right],
$
where $\ell_n(\theta,\mathcal D_n)\equiv\frac{1}{n}\sum_{i=1}^n \ell(\theta, Y_i,\bm X_i)$ and where $w>0$ is referred to as a learning rate. This so-called ``Gibbs posterior"  
corresponds to the actual posterior when $\ell(\theta, Y,\bm X)=-\log\pi(Y\C\bm X, \theta)$. Similarly as in the likelihood-driven power AI priors from Section \eqref{sec:Par}, we can incorporate $\mathcal D_m^*$ through  
\begin{equation}
\pi_{GP}(\theta)\propto \pi_W(\theta)\exp\left[- \alpha\, \ell_m(\theta,\mathcal D_m^*)\right]\label{eq:Gibbs_prior}
\end{equation}
for some working prior $\pi_W(\cdot)$ and a learning rate $\alpha>0$.  Similarly as in Section \ref{sec:EP}, one could consider an expected Gibbs posterior prior version where 
the data $\mathcal D_m^*$ is marginalized out.
Due to the coherency, the Gibbs posterior under the ``Gibbs prior" \eqref{eq:Gibbs_prior} writes as
\begin{equation}
\pi_{GP}(\theta\C \mathcal D_n,\mathcal D_m^*)\propto \pi_W(\theta)\exp\left[- w\, \ell_n(\theta,\mathcal D_n)-\alpha\, \ell_m(\theta,\mathcal D_m^*)\right]. \label{eq:Gibbs_posterior}
\end{equation}
When $r=n/m$ for some $r>0$, i.e. $m\rightarrow\infty$ as $n\rightarrow\infty$,  
 and under suitable regularity conditions on $\ell(\cdot)$ (see  e.g. \cite{Chernozhukov2003} or supplemental material of \cite{lyddon2019general}) the Gibbs posterior has the following asymptotic normal distribution  as $n\rightarrow\infty$
$$
\sqrt{n(1+1/r)}\left(\theta-\wh{\theta}_{n,m}^\alpha\right)\rightarrow z\sim \mathcal N(0,[w J_1(\theta_0)+\alpha J_2(\theta_0)]^{-1}), 
$$ 
where 
\begin{equation}
J_1(\theta )=\int\nabla^2\ell(\theta, Y ,\bm X )\d F_0[(Y,\bm X)]\quad\text{and}\quad J_2(\theta)=\int\nabla^2\ell(\theta, Y ,\bm X )\d F_{AI}[(Y,\bm X)]\label{eq:Js}
\end{equation}
where 
\begin{equation}
\wh\theta_{n,m}^\alpha=\arg\min\{w\, \ell_n(\theta,\mathcal D_n)+\alpha\, \ell_m(\theta,\mathcal D_m^*)\}\label{eq:biased_estimate}
\end{equation} 
is a potentially biased estimator of $\theta_0$. This result can be shown by simple adaptation of general Gibbs posterior theory developed earlier in \cite{Chernozhukov2003}.
While the Gibbs posterior \eqref{eq:Gibbs_posterior}  can be a useful inferential object, simulating from it using MCMC can be at least as challenging as simulating from regular posteriors (see \cite{Nie2022} and references therein). We consider a related, but computationally far more feasible, strategy
 that performs simulation through optimization of randomized objectives. Such strategies have proven useful in various contexts including high-dimensional variable selection \citep{Nie2023}.

 \subsection{Non-parametric AI Priors}\label{sec:NP_AI_priors}
 Rather than expressing prior beliefs about $\theta_0$ defined in \eqref{parameter_loss} through  $\pi(\cdot)$, we can express them through a prior $\pi(F)$ on $F_0$. This will lead to a procedure that is related conceptually but computationally  quite different compared to MCMC sampling from \eqref{eq:Gibbs_posterior}. 
Since the sampling distribution $F_0$ is unknown, we can place a Dirichlet process  (DP) prior with an AI base prior as follows
\begin{equation}
F\sim DP(\alpha, F_{AI}),\label{eq:DP_prior}
\end{equation}
where $\alpha>0$ is the usual concentration parameter and $F_{AI}$ is the base measure which can only be accessed though its  simulations $(Y_i^*,\bm X_i^*)$. 
\subsubsection{AI Base Measure}\label{subsec:AI_base}
Denote the density of this base distribution
as $f_{AI}(Y^*, \bm X^*)$  and factorize it into 
$$
f_{AI}(Y^*, \bm X^*)= f_{AI}^X(\bm X^*)\times f_{AI}^{Y}(Y^*\C \bm X^*).
$$ 
The density $f_{AI}^X(\bm X^*)$ can be viewed as a  distribution over prompts. 
For our practical illustrations, we will assume that it is based on the observed covariates, i.e. $f_{AI}^X(\bm X^*)=\sum_{i=1}^ng_i\delta_{\bm X_i}$ for some (fixed or random) weights $g_i>0$ such that $\sum_{i=1}^ng_i=1$.  Given the prompt $\bm X^*$, the density $f_{AI}^{Y}(Y^*\C \bm X^*)$ is defined implicitly by the AI generator $\hat\mu(\bm X^*)$, be it ChatGPT or any other black-box predictive model.  Perhaps the simplest way to construct $f_{AI}^{Y}(Y^*\C \bm X^*)$ would be an empirical distribution of this historical data, i.e.  $f_{AI}^{Y}(Y^*\C \bm X^*)=\frac{1}{m}\sum_{j=1}^m\delta_{(Y_j^*,\bm X_j^*)}$, where  $\mathcal D_m^*=\{(Y_j^*,\bm X_j^*)\}_{j=1}^m$ have been generated hierarchically from $\bm X_j^*\sim f_{AI}^X$ and then $Y_j^*= \hat\mu(\bm X_j^*)$. Using the log-likelihood loss function, this strategy is closely related to  the power priors discussed in Section \ref{sec:Par} that treat the historical observations as fixed. Just like with posterior-expected priors  from Section \ref{sec:EP}, however,  it might be desirable to incorporate randomness in $\mathcal D_m^*$ and treat $f_{AI}$ (or at least $f_{AI}^{Y}$) as a continuous density.   From the properties of the DP prior \citep{van1998asymptotic},  as $n\rightarrow\infty$ asymptotic consistency of the posterior \eqref{eq:DP_posterior} is achieved under certain regularity conditions regardless of the choice of $F_{AI}$ \citep{lyddon2019general}.

\subsubsection{Posterior Bootstrap}
 The prior distribution on $\theta$ is implied by a prior distribution on $F$  in \eqref{eq:DP_prior} using the mapping \eqref{parameter_loss} where
 $$
 \theta\sim \argmin_{\theta'}\int \ell(\theta',Y,\bm X)\d F [(Y,\bm X)]\quad\text{for}\quad F\sim DP(\alpha, F_{AI}).
 $$ 
 From the conjugacy of the DP process, we see that having observed $\mathcal D_n$, the posterior on $F$ satisfies
$F\C \mathcal D_n\sim DP(\alpha+n, G_n)$ where $G_n=\frac{\alpha}{\alpha+n}F_{AI}+\frac{1}{\alpha+n}\sum_{i=1}^n\delta_{Y_i,\bm X_i}.$
The non-parametric posterior on $\theta$ can be then computed \cite{fong2019scalable}  simply by taking a functional of samples $F$ from its posterior using
\begin{equation}
 \theta\sim \argmin_{\theta'}\int \ell(\theta',Y,\bm X)\d F [(Y,\bm X)]\quad\text{for}\quad F\sim DP(\alpha+n, G_n).\label{eq:DP_posterior}
 \end{equation}
With an empirical AI base measure based on historical data $\mathcal D_m^*$,  the $t^{th}$ posterior sample $\theta^{(t)}$ can be simply computed as
\begin{equation}
\theta^{(t)}=\arg\min_{\theta'}\left[\sum_{i=1}^{n} w_j^{(t)} \ell(\theta,Y_i,\bm X_i)+\sum_{j=1}^{m} w_j^{*(t)} \ell(\theta,Y^*_j,\bm X^*_j) \right]\label{eq:DP_posterior_finite}
\end{equation}
 where $w_j^{(t)}$ and $w_j^{*(t)}$  are DP-posterior implied weights whose refreshment  induces a posterior for $\theta$. With a continuous base prior $f_{AI}^{Y}$  for the labels $Y^*$, the second sum in \eqref{eq:DP_posterior_finite} is infinite and approximate computation is required. One possibility is to perform approximate sampling from the DP posterior using the Posterior Bootstrap Algorithm  (Algorithm 2 in \cite{fong2019scalable} outlined in Algorithm~\ref{alg:PB}). {The idea is to first sample $m$ i.i.d observations $\mathcal{D}_{m}^{*}$ from the base measure $f_{AI}$ for some truncation size $m\in\mathbb{N}$. Then, we assign a random weight to each observation in $\mathcal D_n$ and $\mathcal D_m^*$  perform repeated optimization of the randomized objective through \eqref{eq:DP_posterior_finite}. Note that $m$ does not correspond to the ``strength'' of the prior, but only to the number of atoms in the atomic distribution used to approximate the base measure. Indeed, $\alpha$ is the parameter that corresponds to the strength of the prior, with the intuition that the prior is ``as strong as'' the likelihood of $\alpha$ data points.} The posterior distribution is induced by uncertainty in $F$  and, since $\theta$  is a functional of $F$, we can obtain posterior distribution for a wider class of parameters $\theta$ than possible within a classical likelihood-based Bayesian analysis \cite{chamberlain1996nonparametric}. This  could be viewed as one possible  Bayesian approach to M-estimation.

\begin{algorithm}[!t]
\caption{Posterior Bootstrap}\label{alg:PB}
\begin{algorithmic}[1]
\Require Input observed data $\mathcal D_n$, concentration $\alpha>0$ and approximation truncation  $m$.
\For{$t \gets 1$ to $B$}
    \State Draw imaginary data $\mathcal D_m^*=\{(Y_i,\bm X_i^*)\}_{i=1}^m$ from $f_{AI}$ defined in Section \ref{subsec:AI_base}.
    \State Draw weights $(w_{1:n}^{(t)},w_{1:m}^{*(t)} )$ from $\mathrm{Dir}(1,\dots, 1,\alpha/m,\dots, \alpha/m)$.
    \State Compute $\theta^{(t)}$ from \eqref{eq:DP_posterior_finite}.
\EndFor
\State \Return Posterior Bootstrap sample $\{\theta^{(t)}\}_{t=1}^B$.
\end{algorithmic}
\end{algorithm}

Having obtained samples  $\{\theta^{(t)}\}_{t=1}^B$ through optimization over a dataset consisting of observed and fake labeled data, we can 
proceed with inference (uncertainty quantification) on $\theta_0$ defined in \eqref{parameter_loss} or  posterior predictive inference as follows.
For a likelihood-based loss function $\ell(\theta, Y,\bm X)=-\log\pi(Y\C\bm X, \theta)$, the predictive distribution for $Y_{new}$ given $X_{new}$ could be computed
 though \eqref{eq:predictive_inference} as
 $$
 \pi(Y_{new}\C \bm X_{new})=\frac{1}{B}\sum_{t=1}^B \pi(Y\C\bm X, \theta^{(t)}).
 $$
For example, predicting $Y_{new}\in\{1,\dots C\}$ from $\bm X_{new}$ using a deep learning (DL) classification model with class probabilities $f_\theta(\cdot) \equiv [f_\theta^1(\cdot),\dots, f_\theta^C(\cdot)]$ parametrized by DL weights  $\theta$, we could obtain the non-parametric posterior for $f_\theta$ under the AI prior  using Posterior Bootstrap. The predictive distribution of 
$P[Y_{new}=c\C \bm X_{new},\mathcal D_n]$ for the new label would then be the posterior-averaged class probability $\frac{1}{B}\sum_{t=1}^B f_{\theta^{(t)}}^c(\bm X_{new}).$ The practical utility of the posterior bootstrap for inference on $\theta_0$ can be gauged from its asymptotic distribution.

\section{Theory}\label{sec:theory}

Consider data $Y_{1},\ldots,Y_{n}\sim F_{0}$ and a base measure $F_{AI}$. If $F_{AI}$ is a mixture of $m$ point masses, we denote these by $Y^{*}_{1},\ldots,Y^{*}_{m}$. Otherwise, we let $Y^{*}_{1},\ldots,Y^{*}_{m}\sim F_{AI}$ for some suitably large $m$ as in the posterior bootstrap algorithm. { Recall that the DP prior parameter $\alpha$ can be interpreted as the effective sample size contribution of the prior. Having a constant $\alpha$ independent of $n$ will cause the prior to become irrelevant in the asymptotic regime $n\to\infty$. For this reason, we let $\alpha$ depend on $n$ and fix $\alpha=\gamma n$ for some constant $\gamma>0$. This can be interpreted as fixing the proportion of contribution of total effective sample size that comes from the AI prior}. We define the oracle risk minimizer
\begin{equation}
  \theta_{0}^{\gamma} = \argmin_{\theta}\left[\int \ell(\theta,Y)\,\d F_{0}(Y) + \gamma \int \ell(\theta, Y)\,\d F_{AI}(Y)\right]\,.
  \label{eq:oracle-riskmin}
\end{equation}
noting its dependence on $\gamma$. Similarly, we define the empirical risk minimizer
\begin{equation}
  \wh{\theta}^{\alpha}_{n} = \argmin_{\theta}\left[ \frac{1}{n}\sum_{i=1}^{n} \ell(\theta,Y_{i}) +\frac{\gamma}{m}\sum_{j=1}^{m} \ell(\theta,Y^{*}_{j})\right]\,.
  \label{eq:empirical-riskmin}
\end{equation}

% Define the generalized information matrix $I:\Theta \to \R^{p\times p}$ by
% \[
% (1+\gamma)I(\theta) = I_{1}(\theta) + \gamma I_{2}(\theta)
% \]
% where
% \[
% I_{1}(\theta) = \int \nabla \ell(\theta,Y) \nabla \ell(\theta,Y)^{\top} \,\d F_{0}(Y)\,,\qquad I_{2}(\theta) = \int \nabla \ell(\theta,Y) \nabla \ell(\theta,Y)^{\top} \,\d F_{AI}(Y).
% \]

\begin{theorem}\label{thm:PB-appendix}
\sloppy Let $\theta^{*}$ be the posterior bootstrap sample  obtained from Algorithm~\ref{alg:PB} and denote with $\Pi_{PB}$ its probability measure. Consider the base measure $F_{AI}$ to be atomic with $m$ atoms. Under regularity conditions, for any Borel set $A\subset\Theta\subseteq \R^d$ with $\alpha=\gamma n$ as $n\rightarrow\infty$ we have
$$
\Pi_{PB}\left[\sqrt{n(1+\gamma)} \left(\theta^*-\wh\theta_{n}^\alpha\right)\in A\right]\rightarrow P(z\in A)
$$
$\mathcal D_n$-almost surely   where $\wh\theta_{n}^\alpha$ is the empirical risk minimizer \eqref{eq:empirical-riskmin} and where $z\sim N\left(0, J(\theta_0^{\gamma})^{-1}I(\theta_0^{\gamma})J(\theta_0^{\gamma})^{-1}\right)$ with $(1+\gamma)J(\theta)=J_1(\theta)+\gamma J_2(\theta)$ and $(1+\gamma)I(\theta)=I_1(\theta)+\gamma I_2(\theta)$ where  (denoting $\nabla$ the gradient operator with respect to $\theta$)
\begin{equation}
J_1(\theta )=\int\nabla^{2} \ell(\theta, Y )\d F_0(Y) \,,\qquad J_2(\theta)=\int\nabla^{2}\ell(\theta, Y )\d F_{AI}(Y).\label{eq:Js}
\end{equation}
and
\begin{equation}
I_1(\theta )=\int\nabla \ell(\theta, Y )\nabla \ell(\theta, Y )^T\d F_0(Y) \,,\qquad I_2(\theta)=\int\nabla\ell(\theta, Y )\nabla \ell(\theta, Y )^T\d F_{AI}(Y).\label{eq:Is}
\end{equation}
{
When the base measure is continuous, an analogous result holds where the posterior bootstrap algorithm employs a truncation size $m$ that grows with $n$ and satisfies  $m/n \to r$ for some constant $r>0$.}
\end{theorem}

% prev - before modification
% \begin{theorem}\label{thm:PB}
% Let $\theta^{*}$ be the posterior bootstrap sample  obtained from Table \ref{alg:PB} and denote with $\Pi_{PB}$ its probability measure. Under regularity conditions, for any Borel set $A\subset\Theta\subseteq \R^d$ with $n/m=r$ as $n\rightarrow\infty$ we have
% $$
% \Pi_{PB}\left[\sqrt{n(1+1/r)} \left(\theta^*-\wh\theta_{n,m}^\alpha\right)\in A\right]\rightarrow P(z\in A)
% $$
% $\{\mathcal D_n,\mathcal D_m^*\}$-almost surely   where $\wh\theta_{n,m}^\alpha$ is the empirical risk minimizer~\eqref{eq:biased_estimate} with $\omega=1$ and where $z\sim N\left(0, J(\theta_0)^{-1}I(\theta_0)J(\theta_0)^{-1}\right)$ with $J(\theta)=J_1(\theta)+\alpha J_2(\theta)$ as defined in \eqref{eq:Js} and $I(\theta)=I_1(\theta)+\alpha I_2(\theta)$ where  (denoting $\nabla$ the gradient operator with with respect to $\theta$)
% \begin{equation}
% I_1(\theta )=\int\nabla \ell(\theta, \cdot )\nabla \ell(\theta, \cdot )^T\d F_0 \quad\text{and}\quad J_2(\theta)=\int\nabla\ell(\theta, \cdot )\nabla \ell(\theta, \cdot )^T\d F_{AI}.\label{eq:Js}
% \end{equation}
% \end{theorem}

\proof{Appendix~(See Section \ref{sec:proof}). The result follow the same process as Theorem 1 in \cite{lyddon2019general} whose proof hinges on Theorem 7 and Chapter 3 in \cite{Newton1991} where all the regularity conditions are stated.}
% {\color{red} Sean, can you please verify that this theorem is correct? Also the Gibbs posterior limiting distribution I derived earlier. It might be nice to have a formal proof if you could write it down. It might not be necessary though if it is somewhat obvious.} {\color{purple} In appendix}

It it curious to compare the asymptotic distribution of posterior bootstrap and the Gibbs posterior \eqref{eq:Gibbs_posterior}. As noted earlier by \cite{bissiri2016general} in the context of loss-likelihood bootstrap, the centering is the same but the covariance matrices are different when the loss function is not the usual log-likelihood in which case the bootstrap supplies the usual sandwich covariance matrix.  
The centering $\wh\theta_{n}^\alpha$ of the asymptotic distribution may be a biased estimator of $\theta_0$ when the prior influence does not vanish as $n\rightarrow \infty $ (i.e. when $\alpha=\gamma n$ and $\gamma\geq 0$) and when the AI algorithm provides predictions that are systematically biased.  Indeed, while in the parametric Bayesian framework the prior influence  vanishes as $n\rightarrow\infty$, in Theorem \ref{thm:PB-appendix} we allow for $\alpha\rightarrow\infty$ yielding a possibly biased centering with the amount of bias determined by $\gamma>0$. The prediction-powered inference framework \citep{angelopoulos2023prediction,Angelopoulos2023} estimates the severity of the  bias on the labeled data $\mathcal D_n$ by comparing observed labels $Y_i$ to the AI-predicted ones $\wh\mu(\bm X_i)$. We could apply a de-biasing strategy similar to theirs in order to obtain a centering that is unbiased for $\theta_0$.

\subsection{The Concentration Parameter}\label{sec:alpha}
The concentration parameter $\alpha>0$ measures the assuredness of the prior about $F_{AI}$ which can be interpreted as the effective sample size of the imaginary data $\mathcal D_m^*$.
This can be seen from the characterization of the posterior in \eqref{eq:DP_posterior}. While $m$ is the actual sample size for $\mathcal D_m^*$, we treat it more as a truncation parameter in an approximation to the DP posterior where (similarly as for the catalytic priors in Section \ref{sec:catalytic}) the larger $m$ is, the better. We can choose $\alpha$ adaptively from out-of-sample experiments to determine the relevance of the AI non-parametric prior for prediction and to find the most suitable degree of AI prior subjectivity.

\subsubsection{Calibration via Coverage}
{Another option is to choose $\alpha$ in order to calibrate the coverage of posterior credible intervals in the frequentist sense. One way to do this would be via an adaptation of the general posterior calibration algorithm of \citep{syring2018calibration}. Given access to mechanism to repeatedly accrue samples of size $n$ from the data-generating process, as well as knowledge of a true parameter of interest $\theta^{*}$, one would be able to choose $\alpha$ such that a $1-\delta$ posterior credible interval arising from the $\alpha$-AI prior has frequentist coverage at level $1-\delta$. This would be approximated by repeatedly sampling datasets $\mathcal{D}_{n}$, computing the $1-\delta$ credible region, and determining the proportion of times that the true parameter lies within. The practitioner would then want to choose the largest value of $\alpha$ such that the $1-\delta$ credible interval has frequentist coverage at level $1-\delta$, maximizing the informativeness of the prior under the constraint of well-calibrated posterior credible regions. Of course, practitioners do not have access to $\theta^{*}$ or the ability to generate new data samples. Adapting the general posterior calibration algorithm~\citep{syring2018calibration}, we can replace sampling independent datasets with bootstrapping datasets of size $n$ from the empirical distribution of the actual sample $\mathcal{D}_{n}$. Similarly, knowledge of $\theta^{*}$ is replaced with the empirical risk minimizer on the bootstrapped dataset. One can then solve for $\alpha$ using the same criterion-- the largest value of $\alpha$ such that the estimate of the coverage arising from the bootstrapped samples is at least $1-\delta$.
}

\subsubsection{Asymptotic Calibration}

Adaptive tuning has been also considered in the context of prediction-powered inference by \cite{Angelopoulos2023} who consider a weighted average of loss functions for estimating $\theta_0$ with the weight chosen adaptively from data to minimize the Fisher information number, i.e. trace of the inverse  asymptotic covariance matrix. This weight calibration is related to the one considered in \cite{bissiri2016general} who calibrate a weight  of the Gibbs posterior by matching the asymptotic covariance matrices of the Gibbs posterior and the loss-likelihood bootstrap. 
While these calibrations are ultimately asymptotic as $n\rightarrow\infty$, we could consider a similar strategy to find $\alpha$ that calibrates the trace of an estimate of $J(\theta_0)^{-1}$ in Theorem~\ref{thm:PB-appendix}. This estimate could be obtained by replacing $J_0(\cdot)$ and $I_0(\cdot)$ with their finite-sample counterparts and replacing $\theta_0$ with $\wh \theta_n$. Defining $\hat{\Sigma}(\alpha)$ to be the estimate of the asymptotic covariance, this implicitly defines a function $\alpha\mapsto \hat{\Sigma}(\alpha)$ which can be solved for $\alpha$ when equated to a reasonable target.
In many cases, such a target can be identified from the asymptotic marginal variances used to construct confidence intervals for the PPI estimator \citep{angelopoulos2023prediction}. Indeed, defining $\hat{\sigma}^{2}_{n,N,j}$ to be the $j$th component of the $p$-dimensional parameter of interest, we can solve the equation $\mathrm{tr}\left(\wh{\Sigma}(\alpha)\right)^{-1}=\sum_{j=1}^{p}\hat{\sigma}^{2}_{n,N,j}$ to find a value of $\alpha$ such that the size of the credible intervals are calibrated relative to PPI. In our experiments, we use both of the aforementioned strategies for eliciting values of $\alpha$ for the DP prior, and generally find relatively compatible results (refer to Section~\ref{sec:additional-experiments}).

We can calibrate the DP prior parameter $\alpha$ via the asymptotic covariance in Theorem~\ref{thm:PB-appendix}. Define $\Sigma(\gamma)=J(\theta_{0}^{\gamma})^{-1}I(\theta_{0}^{\gamma})J(\theta_{0}^{\gamma})^{-1}$. In practice, the true risk minimizer $\theta_{0}^{\gamma}$ as well as population quantities $J(\theta_{0}^{\gamma})$,$I(\theta_{0}^{\gamma})$, are not available. We can estimate the asymptotic covariance using the empirical versions of the information matrices and the empirical risk minimizer as
\[
\wh{\Sigma}(\alpha)=J_{n}^{\alpha}(\wh{\theta}_{n}^{\alpha})^{-1}I_{n}^{\alpha}(\wh{\theta}_{n}^{\alpha})J_{n}^{\alpha}(\wh{\theta}_{n}^{\alpha})^{-1}\,.
\]
Consider the problem of mean estimation first. Algorithm 1  of~\citet{angelopoulos2023prediction} provides an asymptotically valid confidence interval for the PPI estimator. In their notation, where $\hat{\sigma}^{2}_{f}$ denotes the empirical variance of the imputed estimate, and $\hat{\sigma}^{2}_{f-Y}$ denotes the empirical variance of the rectifier, the asymptotic variance of the PPI estimator for the mean using $n$ datapoints and $m$ imputed samples has the form $\hat{\sigma}^{2}_{f}/n + \hat{\sigma}^{2}_{f-Y}/m$. We could proceed to calibrate our DP prior parameter $\alpha$ by equating the asymptotic variance of our posterior bootstrap samples with the asymptotic variance of the PPI estimator, by solving the equation
\[
\frac{\mathrm{tr}\,\,\wh{\Sigma}(\alpha)}{n+\alpha} = \frac{\hat{\sigma}^{2}_{f}}{n} + \frac{\hat{\sigma}^{2}_{f-Y}}{m}\,.
\]
This can be solved in terms of $\alpha$ via an iterative root-finding algorithm. This can be extended in a straightforward manner to other estimation problems in which the asymptotic covariance of the PPI estimator is known, such as the linear regression coefficient vector.

For estimands arising from general nondegenerate convex optimization problems (see Algorithm 5 of~\citet{angelopoulos2023prediction}), PPI provides a confidence set in parameter space directly rather than an asymptotically normal point estimator. In this case, we can similarly choose $\alpha$ by matching the width of the AI prior-induced credible interval constructed using the asymptotic covariance from Theorem~\ref{thm:PB-appendix} with that of the PPI confidence set. For multi-dimensional estimands, we choose to match the sum of the axis-aligned extents in each component dimension.

% {\color{red} Sean, I wonder if you could derive this estimator by minimizing the trace of the inverse in Theorem 1. I also think that $\theta_0$ should be replaced by the unbiased estimator $\wh\theta_n$ as opposed to $\wh\theta_{n,m}^\alpha$. Let me know what you think.

% It seems that $\alpha$ should solve something like
% $$
% (1+\alpha g)^2(2 A_{21}+2\alpha A_{22})=A_{11}+A_{21}^C+A_{22}^C
% $$
% where $g=tr(I_2 I_1^{-1})$ and $A_{ij}=tr(J_i I_1^{-1}J_j)$ and $A_{ij}^C=tr(J_iCJ_j)$ with $C=I_1^{-1}I_2 I_1^{-1}$.
% I might be wrong.
% }

\section{Generative AI Illustrations}\label{sec:Applications}
We demonstrate our approach on two classification datasets, where generative AI output could be incorporated in predictive inference for medical diagnosis or parameter inference in labeling massive galaxy images. We also discuss calibration of $\alpha$ in a gene expression example.

\subsection{Skin Disease Prediction}
%We use multi-category random forests for predictive inference. This means that we will integrate predictions out-of-sample over the bootstrap posterior obtained on combined observed and fake data.

We apply our methodology towards the classification of Erythemato-Squamous diseases from descriptions of clinical symptoms. Erythemato-Squamous diseases (ESDs) comprise a group of six distinct but closely overlapping skin conditions that pose significant diagnostic challenges due to their similar clinical and histopathological features. Machine learning approaches have been applied to predict the disease subtype from these clinical and histopathological features with high accuracy, additionally providing interpretable patterns \citep{wang2025identifying}. This dataset has also been employed for exploring uncertainty quantification in large language model-based medical diagnosis. \citet{kim2024adaptive} used ChatGPT to diagnose ESDs from descriptions of clinical features only, applying conformal prediction techniques to aid in uncertainty quantification. Notably, ChatGPT's diagnoses from clinical symptoms only were less accurate than that of bespoke machine learning algorithms (i.e. a simple random forest model, for example), but still substantially better than random guessing.

ESDs are divided into the following six subtypes, which are the labels in this classification problem: {\em psoriasis, seborrheic dermatitis, lichen planus, pityriasis rosea, chronic dermatitis and pityriasis rubra pilaris}. There are twelve clinical features, ten of which are ordinal variables that take values in the set $\{0,1,2,3\}$, describing levels of prevalence. These features include things like redness, itchiness, or incidence in certain regions of the body. The other variables are family history (binary) and age (integer valued).  {Clinical features are those based on observable signs and symptoms that can be identified through physical examination and patient history, in contrast to histopathological features which are typically examined under a microscope after a biopsy. %We denote the feature space by $\mathcal{X}$.
}

Through our framework of generative AI priors, we leverage pre-trained large language models (ChatGPT) as complementary diagnostic tools to enhance the predictive capabilities of traditional machine learning systems at classifying skin disease diagnoses correctly.

\subsubsection{Data}

We analyze the dermatology data\footnote{Available at \url{https://archive.ics.uci.edu/dataset/33/dermatology}} available from the UCI machine learning repository \citep{dermatology_33}, removing histopathological features so as only to diagnose disease from clinical features. For this experiment, we split the total number of observations in the dataset (366) as follows: 15\% is used as training data, treated as correctly labeled pairs. 20\% is held-out to assess test accuracy. The remaining 65\% is considered to be extra unlabeled data, for which the clinical symptoms are known to the practitioner but the labels are not. We let $\mathcal D_n$ denote the labeled training data, $  {\mathcal D}^{Test}_{T}$ the labeled testing data and denote the extra unlabeled data as $\mathcal D_m^*$. We conduct our experiments ten times under different random splits of training, imputation, and test data.
%For each unlabeled observation $\bm X^*_{i}$ for $i=1,\ldots,m$, we impute $R$ labels $\tilde{Y}_{i}^{(1)},\ldots,\tilde{Y}_{i}^{(R)}$ by sampling from the distribution on labels induced by querying ChatGPT 4o to assign a diagnosis to this set of features. {\color{red} this is different than using $m$ random samples from empirical X and then querying. Can we try this approach instead?} Yes -- i'm updating accordingly

\subsubsection{Prompting ChatGPT to Impute Diagnoses}\label{subsec:prompting}
As discussed in Section \ref{subsec:AI_base}, the base measure $F_{AI}$ for our AI prior is characterized by a probability distribution on both clinical features $\bm X$ and labels $Y$. In this case, we define such a base measure as follows: the marginal distribution of clinical features is from the empirical distribution of extra unlabeled data, that is, $f_{AI}^X(\bm X^*)=\frac{1}{n}\sum_{i=1}^n \delta_{\bm X_i}(\bm X^*)$. {Then, the conditional AI prior on the labels is given by the GPT-imputed conditional distribution on the feature $\bm{X}$ using the strategy  described below.

In order to convert use ChatGPT to predict labels (diagnoses) from clinical features, we first convert the set of features to a prompt. We elicit these responses from the \texttt{o4-mini} language model, using a temperature setting of 0.7. For this experiment, we used prompts with the following general format\footnote{The prompt shown here is slightly simplified, excluding instructions on how to format the output. The prompt shown here is simplified and excludes detailed instructions. The exact verbatim prompt is provided in Section~\ref{sec:prompt}. }

\prompt{
\textbf{Prompt:} Predict the diagnosis of Erythemato-Squamous disease from the following clinical features. The age feature simply represents the age of
    the patient. Family history is a binary variable. Every other feature was given a degree in the range of 0 to 3. Here, 0 indicates that the feature was not present, 3 indicates the largest amount possible, and 1, 2 indicate the relative intermediate values.

    erythema: 2, scaling: 2, itching: 3, [...], age: 55

	[...] Estimate the probability of each possible diagnosis for this case [...]

\texttt{o4-mini} psoriasis: 0.45, lichen planus: 0.20,  [...]
}

We parse the textual response into a probability distribution on classes using regular expressions. We employ an additional normalizing step to ensure the given probabilities sum to one, though it is almost never necessary. We take the class with largest probability as the AI-predicted label, breaking ties uniformly at random if necessary.

% the empirical distribution of the $R$ GPT-imputed labels {\color{red} I think you should do m random samples not R samples for each i} that correspond to the given clinical feature value.

\begin{figure}
    \centering
	\includegraphics[width=0.70\textwidth]{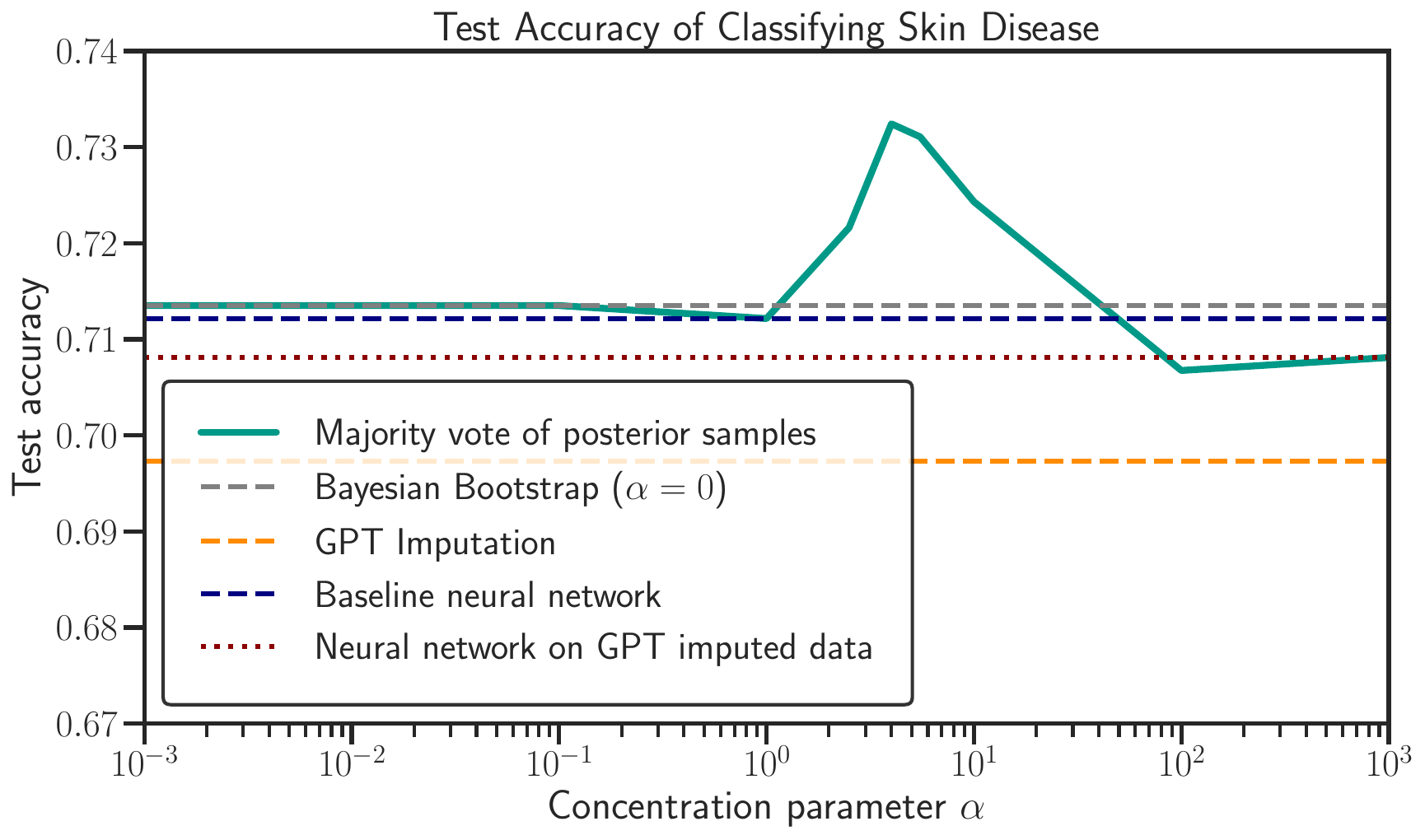}
    \caption{Classification accuracy of ESD on held-out test data, using a neural network trained on $n=58$ observations. Each line indicates the mean performance after 10 repetitions. Horizontal lines indicate the test performance of a fitted neural network fit on training data only, of ChatGPTs imputations, and of a neural networked fit only on imputed data. }
    \label{fig:skin-disease-1}
\end{figure}

\subsubsection{Non-parametric AI Bayesian Inference}\label{subsubsec:np-ai-bayes}
For this data, we posit a parametric model for the conditional probabilities of each label via a three-layer neural-network parameterized vector function $f_\theta:\mathcal{X}\to \mathcal{S}^{6}$, where $\mathcal{S}^{6}:=\{v\in \mathbb{R}^{6}:\sum_{i=1}^{6}v_{i}=1,\,v_{i}\geq 0 \,\forall i=1,\ldots,6\}$ denotes the simplex on $6$ elements. {The architecture of the neural network is rather uncomplicated and is described below.

The neural network parameterizes a class of functions $f_{\theta}:\mathcal{X}\to \mathcal{S}^{K}$ through the following function composition
\[
f_{\theta} = \mathrm{softmax}\circ f_{\bm{W}_{2},\bm{b}_{2}}\circ \sigma \circ f_{\bm{W}_{1},\bm{b}_{1}}
\]
where $\theta=(\bm{W}_{1},\bm{b}_{1},\bm{W}_{2},\bm{b}_{2})'$, and $\mathbf{W}_1 \in \mathbb{R}^{20 \times 12}, \mathbf{b}_1 \in \mathbb{R}^{20}$, $\mathbf{W}_2 \in \mathbb{R}^{6 \times 20}, \mathbf{b}_2 \in \mathbb{R}^{20}$. In addition, $\sigma$ denotes the ReLU activation function $\sigma(x)=\max\{0,x\}$, $f_{W,b}$ denotes the affine transformation $f_{\bm{W},\bm{b}}(x)=\bm{W}x+\bm{b}$, and $\mathrm{softmax}$ is defined via
\[
\text{softmax}(\mathbf{z})_i = \frac{\exp(z_i)}{\sum_{j=1}^6 \exp(z_j)}\,.
\]
The neural network parameters are fit using the Adam optimizer to minimize the weighted cross-entropy loss, with a learning rate of 0.001. The clinical symptom covariates are first preprocessed by standardizing the age feature (by subtracting its mean and dividing by the unbiased estimate of its standard deviation).

  Our inferential target is the minimizer of the induced empirical classification loss on the neural network weights $\theta$. The Posterior Bootstrap  distribution  $\{\theta^{(t)}\}_{t=1}^B$ obtained from Algorithm~\ref{alg:PB} induces a posterior distribution on $f_\theta(\cdot)$ and thereby also posterior predictive distribution on the label $Y_{j}$ corresponding to test data $\bm X_{j}\in \mathcal D_{T}^{Test}$ for $1\leq j\leq T$. The final prediction of the label will be the majority vote
\begin{equation}
\widehat Y_{j}=\arg\max\limits_{c\in \{1,\dots, 6\}} \frac{1}{B}\sum_{t=1}^Bf_{\theta^{(t)}}^c(\bm X_{j})\label{eq:class}
\end{equation}
where $f_{\theta^{(t)}}^c(\cdot)=P[Y=1\C \theta^{(t)},\cdot]$.  We evaluate the performance of this classification rule through the estimated misclassification rate $P[Y_{j}\neq \widehat Y_{j}]$ from $T$ out-of-sample observations.  

We approximately sample $B=100$ samples from the posterior predictive distribution on the label of each held-out test observation using the posterior bootstrap algorithm, using a truncation size of $m=300$. This essentially materializes as the following: for each posterior sample $t=1,\ldots,B$, we fit the neural network using a weighted loss induced by the AI prior, and classify each test point using \eqref{eq:class}.  We fit the neural network using the Adam optimizer with a finite maximum number of epochs, which we note adds an additional degree of approximation to the posterior sampling procedure.

We employ this procedure for a range of $\alpha$ values (the concentration parameter in the Dirichlet Process AI prior discussed in Section~\ref{sec:alpha}), and repeat it for ten repetitions. Figure~\ref{fig:skin-disease-1} displays the average out-of-sample classification accuracy as a function of the concentration parameter $\alpha$. The blue dashed horizontal line indicates the  classification accuracy of a classification rule $\widehat Y_{j, DL}=\arg\max\limits_{c\in \{1,\dots, 6\}} f_{\hat \theta}^c(\bm X_{j})$, where $\hat\theta$ has been estimated purely on the training data. {Due to stochasticity in predictions from the optimization procedure (see Section~\ref{subsubsec:np-ai-bayes}), the line indicates the average accuracy over ten such estimations of $\hat\theta$.}   
% The classification accuracy has been averaged over   ten repetitions of estimating the parameters { {\color{red} why did you resample the training data? It should be a deterministic prediction} to incorporate stochasticity in fitting the model).
Employing our specified ChatGPT-powered AI prior, we see that a range of relatively small $\alpha$ values leads to posterior predictive distributions that are more accurate than predictions that arise when simply excluding the additional unlabeled data.

Interestingly, we note in this case that for $\alpha>25$, the classification accuracy is diminished by the AI prior. Since $\alpha$ can be interpreted as an effective sample size, this means that when we have more than $\approx 40\%$ fake observations, the performance worsens.  As shown in Figure~\ref{fig:skin-disease-1}, the baseline performance of ChatGPT imputations, {computed as the average accuracy taken over ten repetitions using \texttt{o4-mini} to classify each test point. That is, we repeat the following procedure ten times: impute $\hat{Y}_{j}=\hat{\mu}(\bm{X}_{j})$ for each test point indexed by $j$ and compute the test accuracy.}  This hovers at around 70\% classification accuracy. {The maroon dotted line indicates the accuracy of a neural network fit only on GPT-imputed data, which has an accuracy of around 71\%. The majority vote accuracy of the AI posterior shrinks to this level as $\alpha$ grows large, as this is the limiting case that we expect when the influence of the $n$ data points becomes dominated by the prior (i.e when $\alpha\to\infty$ for a fixed $n$).} The gray dashed line in Figure~\ref{fig:skin-disease-1} indicates the performance for $\alpha=0$, where the procedure boils down to the Bayesian bootstrap (Algorithm 2 in \citep{lyddon2019general}) where we obtain uncertainty quantification based on only $\mathcal D_n$. The best out-of-sample prediction error was achieved for $\alpha=4.0$ which yielded a $2.5\%$ increase in prediction accuracy over a procedure that does not use the fake data (with $\alpha=0$). This increase is notable as the AI predictions themselves are not of significantly high quality for this task.

% {\color{red} Can we try $\alpha=0?$} For $\alpha=0$, this procedure boils down to the Bayesian bootstrap (Algorithm 2 in \citep{lyddon2019general}) where we obtain uncertainty quantification based on only $\mathcal D_n$. {\color{red} Maybe the horizontal line on your plot should be the average misclassification with $\alpha=0$ instead.} Done!

% {\color{red} We should conclude with a statement like....
% The best out-of-sample prediction error was achieved for $\alpha=?$ which yielded $??\%$ increase in prediction accuracy over a procedure that does not use the fake data (with $\alpha$).
% Can you fill in the exact numbers?} Done

\subsection{Proportion of Spiral Galaxies}\label{subsec:proportion-of-spiral}

Prior work has collected human annotations of galaxy morphologies through the Galaxy Zoo 2 citizen science initiative \citep{willett2013galaxy}, which contains over 1.3 million labeled images from the Sloan Digital Sky Survey. \citet{angelopoulos2023prediction} estimate the proportion of galaxies exhibiting spiral arm features, which is useful for understanding stellar evolution and star formation.
The setting is that the practitioner has access to a small number $n\leq 1000$ of human-labeled data (galaxy images pair with human annotations), and a large quantity $N \approx 1.5\times 10^{4}$ of unlabeled galaxy images. A computer vision model is leveraged to impute the labels of these data points. {For the sake of our experiment, we have access to the true labels of the $N$ data points as well, which we additionally use to estimate $\theta^{*}\approx 0.26$ as the ``true mean proportion of spiral galaxies''. However, we use knowledge of $\theta^{*}$ purely for validation, and do not use it nor the true labels of the $N$ computer-vision imputed data points for our analysis.}

We adapt this setting to our AI prior framework, seeking to estimate the proportion of spiral galaxies in the universe. However, rather than using the AI-generated labels on additional data to debias an estimator \citep{angelopoulos2023prediction}, we perform a Bayesian inference on the unknown proportion of spiral galaxies leveraging the AI-predictions to elicit our prior knowledge.

\begin{figure}
    \centering
	\resizebox{0.9\textwidth}{!}{\input{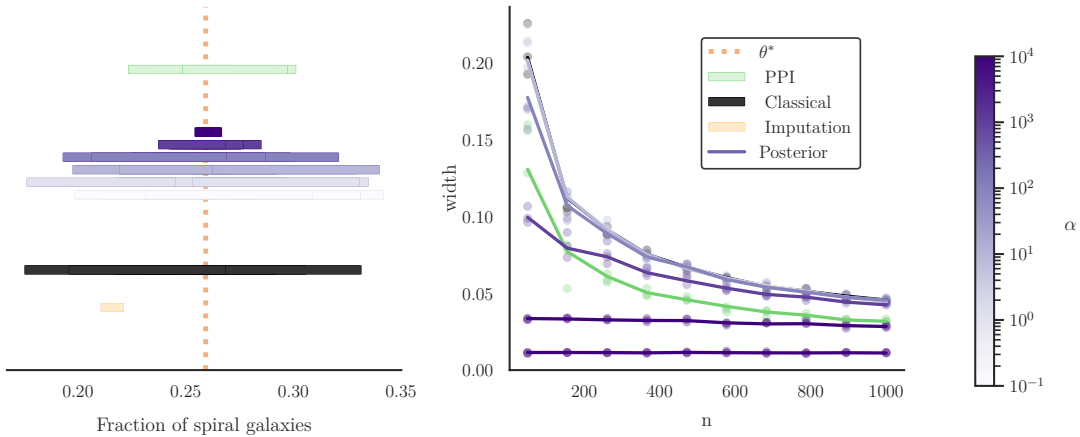}}
    \caption{90\% Credible intervals for estimating the proportion of spiral galaxies. The left plot visualizes a credible interval from our method, compared with 90\% confidence intervals around the classical and PPI estimator. The orange and red bars display confidence intervals around the classical estimator when all imputed data are treated as real, where the imputation is done by thresholding and by sampling respectively. The right plot displays the width of credible/confidence intervals as a function of the labeled training data size $n$. }
    \label{fig:galaxy}
\end{figure}

\subsubsection{AI Priors on the Proportion of Spiral Galaxies}\label{subsubsec:aipriorspiral}

Suppose we have galaxy images $\bm X_{1},\ldots,\bm X_{n}$ with human annotations of their spirality $Y_{1},\ldots,Y_{n}$. Additionally, we have unlabeled galaxy images $\bm{X}^*_{1},\ldots,\bm{X}^*_{N}$ with predicted spirality probabilities ${p}^*_{1},\ldots,{p}^*_{N}$ produced via a large computer vision model. We take the nonparametric approach in Section~\ref{sec:NP_AI_priors} and define an AI base measure $F_{AI}$ that leverages the computer vision model by first sampling an index $j\sim\mathrm{Unif}\{1,\ldots,N\}$ and then sampling a label $Y^{*}\sim\mathrm{Ber}(p^{*}_{j})$.
% This means that we do not have an underlying continuous base measure and can do an exact algorithm without refreshing the labels for a fixed sub-sample of size $m$.
We elicit our AI prior in turn as previously, via $F\sim\mathrm{DP}(\alpha,F_{AI})$ where $F_{AI}$ is this probability measure on $Y^{*}$.  We use the approximate algorithm in Algorithm~\ref{alg:PB} using truncation size $m=10^{5}$, resampling from the empirical distribution of the predicted probabilities and in turn resampling labels from the AI base measure here (the computer vision classifier).}
 
\subsubsection{Nonparametric AI Inference on the Mean}

Our inferential conclusions on the proportion of spiral galaxies stem from the posterior on the risk minimizer $\theta(F)$, defined via $\theta(F)=\argmin_{\theta}\int (y-\theta)\,\mathrm{d}F(y)$. We obtain $B=1000$ samples from the approximate posterior distribution of $\theta (F)$ using the exact variant of the posterior bootstrap algorithm in Algorithm~\ref{alg:PB}. We repeat this procedure 10 times each for various values of the DP concentration parameter $\alpha$ in the AI prior. Figure~\ref{fig:galaxy} displays the sizes of a single 90\% credible interval for varying $\alpha$ values, as well as the size of the 90\% confidence interval for the classical sample mean (using only the human labeled data), the PPI estimator, and the sample mean when treating the imputed labels as real, using two methods of imputation. We note that when labels are imputed by thresholding, there is significant bias that disappears when labels are imputed by sampling. The right hand side of the plot also visualizes the size of these confidence/credible intervals as the sample size $n$ of human labeled data increases.

The posterior distribution on the proportion of spiral galaxies which arises from our AI prior obtains tight 90\% credible intervals that concentrate around the mean. As $\alpha$ grows larger, the predictions from the computer vision model become more heavily incorporated, shrinking the size of the posterior 90\% credible sets. We highlight the power of employing the AI classifier as a base-measure by showing that when a set of imputed data is created by simply thresholding the predicted probabilites above 0.5, the 90\% confidence interval around the sample of these imputed labels is far from the truth as there is nonnegligible bias. Harnessing the AI model's predictive distribution allows the 90\% credible intervals to shrink in size while maintaining desirable frequentist coverage (see Section~\ref{sec:additional-experiments}).

\subsection{Gene Expression and Concentration Parameter Calibration}\label{sec:expression-calib}

\begin{figure}[!t]
    \centering
	\resizebox{0.7\textwidth}{!}{\input{fig/covplot.pgf}}
    \caption{Frequentist coverage of AI posterior 90\% credible intervals for the median expression level of a gene induced by a promoter sequence. Intervals are from the 0.05 to 0.95 posterior quantile, using $n=2000$ samples in the analysis, and the AI prior described in Section~\ref{sec:additional-experiments}. We display the actual coverage computed using oracle knowledge, and a bootstrapped estimated of the coverage computable via sample information only. Vertical lines denote possible choices of $\alpha$.}
    \label{fig:galaxy-coverage}
\end{figure}

The practitioner may also consider wish to consider choosing $\alpha$ such that posterior credible regions are well-calibrated in a frequentist sense. In the spiral galaxy classification task, the predictive distribution of the computer vision classifier is relatively accurate, and allows 90\% credible intervals resulting from AI priors even with very large $\alpha$ to cover the true proportion of spiral galaxies $\theta^{*}\approx 0.26$ (the mean from the entire set of 16,743 labels available, see Section~\ref{subsec:proportion-of-spiral}). We demonstrate two methods for choosing $\alpha$ in another real-world task of inference over the median expression level of a gene induced by a promoter sequence. The AI base measure arises from a transformer model trained to perform this task~\citep{vaishnav2022evolution}. For this task, while the predictive quality of the transformer is still excellent, there will be a turning point where modest values of $\alpha$ cause the bias induced by the AI prior to be too large.

Figure~\ref{fig:galaxy-coverage} displays the frequentist coverage of 90\% posterior credible intervals around the mean (0.05 to 0.95 posterior quantiles). In order to gain the most from the AI prior while maintaining calibration, we can consider choosing the largest $\alpha$ such that the credible interval stays well-calibrated. Credible intervals are constructed using the AI priors constructed using the aforementioned transformer base measure, conditioning on $n=2000$ actual data points. {Actual coverage is calculated from the proportion of intervals containing the true median gene expression level $\theta^{*}\approx 5.65$, which in this case is taken to be the median from the entire set of 61,150 response values available}. The actual coverage (black line) in Figure~\ref{fig:galaxy-coverage} requires the ability to sample new datasets as well as oracle knowledge of the true parameter $\theta^{*}$. The bootstrap estimate of the coverage (orange line) is computed by bootstrapping samples from the empirical distribution of a particular sample of 1000 labeled data points, and computing the proportion of times that the mean of this bootstrapped sample lies inside the interval. This procedure is similar to that in \citep{syring2018calibration} and may be used to approximately calibrate posterior credible regions in absence of the knowledge of the true parameter. In this experiment, the largest $\alpha$ value at which the posterior 90\% credible interval is well-calibrated in the frequentist sense occurs approximately at $\alpha\approx 100$. Asymptotic covariance matching to PPI yields the value $\alpha\approx 1200$, while the bootstrapping calibration algorithm yields the value $\alpha=190$. Note that while in this case the two methods yield values of $\alpha$ that are relatively distant, this is not always the case (see Figure~\ref{fig:appendix-results}). As our Bayesian method embraces bias rather than avoiding it, asymptotic matching to PPI can sometimes result in selecting $\alpha$ too large.

\subsection{Additional Experiments}\label{sec:additional-experiments}

\begin{figure}
    \centering
	\resizebox{0.9\textwidth}{!}{\input{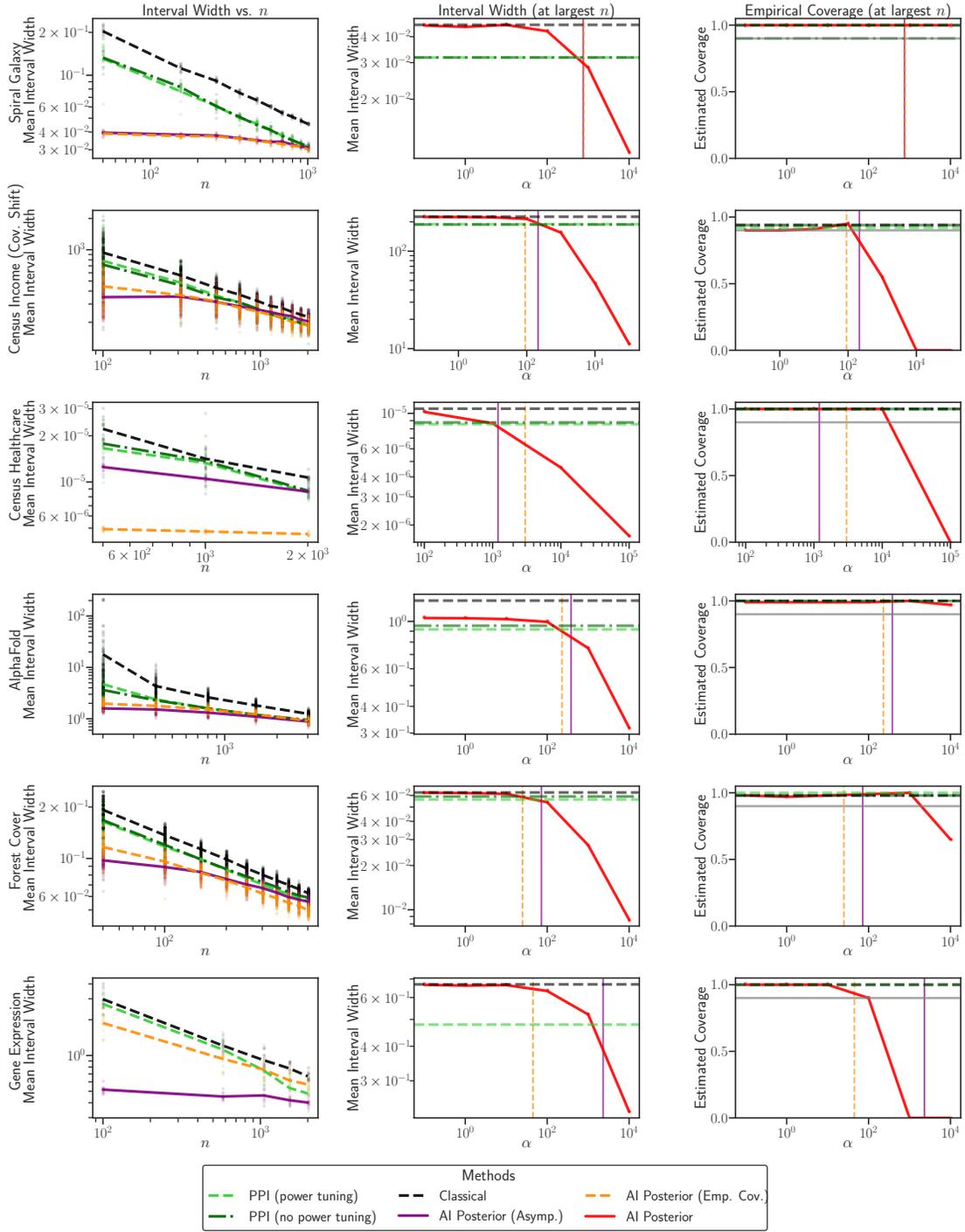}}
    \caption{Interval width and coverage of AI posterior credible intervals on experiments from~\citet{angelopoulos2023prediction}. Left: size of the AI posterior credible interval for specific $\alpha$ choices as a function of $n$. Middle: interval width as a function of $\alpha$ for the largest $n$. Right: empirical coverage of the intervals as a function of $\alpha$ for the largest $n$.}
    \label{fig:appendix-results}
\end{figure}

{We repeat our experimental procedure on six experiments studied in~\citet{angelopoulos2023prediction}. The data was obtained via the \texttt{ppi-python} package. Please refer to this work or the references therein for further information regarding any of these datasets. We provide any relevant details and experimental hyperparameters for each dataset below.

Each experiment follows the setup in prediction-powered inference, in which we have $n$ labeled datapoints, and $N$ unlabeled AI-imputed datapoints. In our case, we refine these $N$ unlabeled datapoints into an AI prior by setting the base measure $F_{AI}$ of our DP prior to be given by their empirical distribution.

For the optimization in each posterior bootstrap iteration, we optimize numerically using the L-BFGS solver if necessary. For each method, we obtain $B=1000$ samples using the posterior bootstrap algorithm. We construct a 90\% credible interval from these set of samples by taking the 0.05 and 0.95 quantiles respectively as endpoints. We repeat each experiment varying over a range of $n$, and $\alpha$ values (where $\alpha$ is the DP prior parameter). We also repeat the experiment for a varying number of repetitions for each combination, so that we can assess variation and compute empirical coverage of the credible intervals. In each repetition, the $n$ labeled datapoints are resampled from a larger body of available data.

Figure~\ref{fig:appendix-results} displays the results regarding 90\% posterior credible intervals for the parameter of interest using AI priors. In the leftmost column, we show the resulting interval widths for two choices of the DP prior parameter $\alpha$. In purple, $\alpha$ was obtained by matching the asymptotic covariance to match that of PPI. In orange, we choose $\alpha$ approximately via the empirical calibration strategy derived from that used for Gibbs posteriors proposed by~\citet{syring2018calibration}. We conclude that across our experiments, leveraging machine learning predictions via AI priors allows us to earn a concentration in posterior mass around the true parameter value. In the second column, we fix the value of $n$ to be the largest that was analyzed for each dataset. We show the size of the 90\% credible interval as a function of $\alpha$. Similarly, in the last column, we visualize the empirical coverage computed as the proportion of repetitions in which the true parameter falls into the interval. We note that this estimation may not be accurate for some of our experiments where only ten repetitions were performed.

\paragraph{AlphaFold.} We construct a credible interval for the odds ratio based on credible intervals for the mean in each group, using the transformation used in~\citep{angelopoulos2023prediction}. We use values $n\in\{200, 400, 800, 1500, 3000\}$ and perform $100$ repetitions.

\paragraph{Census Healthcare.} The parameter of interest is the logistic regression coefficient. We sample from the AI posterior using the posterior bootstrap algorithm minimizing the binary cross-entropy loss. We use values $n\in\{500,1000,2000\}$ and perform $10$ repetitions.

\paragraph{Census Income (covariate shift).} The parameter of interest is the ordinary least squares regression coefficient in the covariate shifted population. We sample from the AI posterior using the posterior bootstrap algorithm minimizing the weighted least-squares loss. We use $n\in\{10,100,2000\}$ and perform 100 repetitions.

\paragraph{Spiral Galaxies.} The parameter of interest is the mean of binary data. This example is described in detail in Section~\ref{sec:Applications}. We use $n\in\{10,50,1000\}$ and perform 10 repetitions.

\paragraph{Forest Cover.} This experiment is carried out exactly as in the spiral galaxy data, as we are interested in the mean of binary data. We use $n\in\{10,50,500\}$ and perform 100 repetitions.

\paragraph{Gene Expression.} The parameter of interest is the 0.5 quantile (also known as the median). We use the posterior bootstrap minimizing the absolute error. We use $n\in\{5,100,2000\}$ and perform 10 repetitions.

}

 \section{Discussion}\label{sec:Discussion}
 This research note proposes a Bayesian alternative to prediction-powered inference  framework  introduced by \cite{angelopoulos2023prediction} for performing valid statistical inference when an experimental dataset is augmented with predictions from an AI system. Our approach is based on prior construction based on simulations from an auxiliary black-box model. Our framework enables uncertainty quantification through non-parametric posteriors by viewing the machine learning system as a simulator from a prior on the unknown distribution function. Treating the generative black-box model as a base measure in the Dirichlet process prior $DP(\alpha, F_{AI})$, we achieve fully Bayesian inference about various quantities of interest (parameters associated with statistical models, parameters defined as minimizers of loss functions) using non-parametric posteriors. These posteriors give rise to posterior predictive distributions in parametric models which can be leveraged for decision making based on both AI input as well as observed data. We estimate the concentration parameter $\alpha\geq 0$ from out-of-sample experiments to determine the inferential usefulness of AI predictions. The estimated value at $\alpha=0$ would signify that AI predictions do not add value and one is better off proceeding without them.
 We find that Bayesian analysis can be meaningfully enhanced with generative AI predictions on two real examples. We found that while AI predictions should not be taken literally for decision making, they can serve as a useful proxy (prior) for the correct answer which could enhance Bayesian analysis of observed data.

\bibliographystyle{chicago}
\bibliography{sources}

\appendix

\section{Appendix}

\subsection{Prompting}\label{sec:prompt}

The exact prompt used for the AI base measure in the skin disease experiment was as follows:

\begin{lstlisting}[breaklines=true, basicstyle=\tiny\ttfamily]
You are an advanced AI medical diagnostic assistant. Your current task is to analyze a set of clinical features related to erythemato-squamous diseases and estimate the probability of six potential diagnoses: psoriasis, seborrheic dermatitis, lichen planus, pityriasis rosea, chronic dermatitis, and pityriasis rubra pilaris.

    Your analysis should be based on a deep understanding of how these features typically manifest in each disease. The input will be a list of features with numerical values. Pay close attention to the scoring system and the typical patterns for each condition.

    Understanding the Input Features:

    The clinical features are scored as follows:

    age: The patient's age in years (continuous variable).
    family_history: A binary variable: 1 indicates a positive family history of relevant dermatological conditions, 0 indicates no such history.
    All other features (erythema, scaling, definite_borders, itching, koebner_phenomenon, polygonal_papules, follicular_papules, oral_mucosal_involvement, knee_and_elbow_involvement, scalp_involvement): These are scored on a scale of 0 to 3.
    0: The feature is absent or not observed.
    1: The feature is present to a mild degree or in a limited extent.
    2: The feature is present to a moderate degree or extent.
    3: The feature is present to a severe degree, is very prominent, or is extensive.
    Detailed Disease Profiles and Feature Interpretation Guidelines:

    Carefully consider the following characteristics for each disease when evaluating the input case. The presence of hallmark features strongly supports a diagnosis, while their absence, or the presence of contradictory features, should lower its probability.

    Psoriasis:
    Key Indicators:
    erythema: Typically bright red (look for high values like 2-3).
    scaling: Classically thick, silvery-white (look for high values like 2-3).
    definite_borders: Lesions are usually sharply demarcated (look for high values like 2-3).
    knee_and_elbow_involvement: Very common sites, particularly extensor surfaces (high values strongly suggestive).
    scalp_involvement: Frequent (moderate to high values are common).
    koebner_phenomenon: Often present (a value > 0 is supportive).
    family_history: Positive family history (1) increases likelihood.
    Less Indicative/Contradictory:
    Ill-defined borders (definite_borders: 0).
    Absence of involvement in classic sites (knees, elbows, scalp) if lesions are present elsewhere.
    Prominent polygonal_papules or follicular_papules are not typical.
    oral_mucosal_involvement is rare.
    Age: Can occur at any age, but common peaks are in the 20s-30s and 50s-60s.
    Seborrheic Dermatitis:
    Key Indicators:
    scaling: Typically greasy, yellowish, or fine white/flaky (moderate values like 1-2).
    erythema: Often pinkish-yellow or mildly red (moderate values like 1-2).
    scalp_involvement: Very common ("dandruff" is a mild form; look for any value > 0, higher if severe).
    Involvement of sebaceous gland-rich areas: face (eyebrows, nasolabial folds, glabella), chest, upper back, flexures (axillae, groin). (The provided features don't specify location beyond scalp/knee/elbow, so infer if possible or weigh scalp heavily).
    Itching: Common, usually mild to moderate (itching: 1-2).
    Definite Borders: Often less defined (definite_borders: 0-1), can be patchy.
    Less Indicative/Contradictory:
    Very thick, silvery scales (more like psoriasis).
    Sharply definite_borders: 3 (less typical than psoriasis).
    Prominent koebner_phenomenon.
    polygonal_papules.
    Significant oral_mucosal_involvement.
    Primary involvement of knees/elbows without scalp or other seborrheic area involvement.
    Age: Common in infants ("cradle cap") and adults (peak 30-60 years).
    Lichen Planus:
    Key Indicators (The "P's"):
    polygonal_papules: Hallmark feature; flat-topped, violaceous (purplish) papules (high values like 2-3 are very suggestive).
    itching: Usually intense (itching: 2-3).
    oral_mucosal_involvement: Common (e.g., Wickham's striae - lacy white pattern); (any value > 0 is significant, higher is more indicative).
    definite_borders: Lesions are typically well-defined (definite_borders: 2-3).
    koebner_phenomenon: Can be present (koebner_phenomenon > 0).
    Erythema: Often has a violaceous (purplish) hue, though the input only gives intensity.
    Typical Locations: Wrists (flexor), ankles, shins, lower back, genitalia. knee_and_elbow_involvement is less classic but possible.
    Less Indicative/Contradictory:
    Absence of polygonal_papules (0) AND absence of oral_mucosal_involvement (0) makes classic Lichen Planus much less likely, even with itching.
    Ill-defined borders.
    Predominantly greasy or silvery scaling (scaling in LP is usually fine or absent on skin papules, though hypertrophic forms can scale).
    Age: Most common in middle age (30-60 years).
    Pityriasis Rosea:
    Key Indicators:
    Often starts with a "herald patch" (not an input feature, but consider if this context was available).
    scaling: Characteristically a "collarette" of fine scales (attached peripherally, loose centrally). (Interpret scaling value with this in mind).
    Distribution: Typically oval, pink-to-tan lesions on the trunk and proximal extremities, often following skin cleavage lines ("Christmas tree" pattern). (The provided features lack distribution detail).
    erythema: Pinkish.
    definite_borders: Individual lesions are fairly well-defined.
    itching: Variable, from absent to severe (itching: 0-3).
    Less Indicative/Contradictory:
    Chronic course (Pityriasis Rosea is usually self-limiting in 6-12 weeks; this is not in the input but is crucial contextual knowledge).
    Predominant involvement of scalp, or isolated knee_and_elbow_involvement.
    Presence of polygonal_papules, thick silvery scales, or follicular_papules.
    Significant oral_mucosal_involvement (rare).
    koebner_phenomenon is rare.
    Age: Most common in older children and young adults (10-35 years). An older age (e.g., >40-50) makes it less typical.
    Chronic Dermatitis (e.g., Atopic Dermatitis, Nummular Eczema):
    This is a broader category. Consider this diagnosis if features are less specific to the other conditions, especially with high itching.
    Key Indicators (General/Atopic):
    itching: Very common and often the dominant symptom, can be severe (itching: 2-3).
    erythema and scaling: Can be variable, skin often dry, may become lichenified (thickened from chronic scratching) with chronicity.
    definite_borders: Often ill-defined (definite_borders: 0-1), especially in atopic dermatitis. (Nummular eczema, a subtype, has well-defined, coin-shaped lesions).
    family_history: For atopic dermatitis, a positive family history (1) of atopy (eczema, asthma, hay fever) is common.
    Location (Atopic): Flexural areas (elbow and knee creases), face, neck. knee_and_elbow_involvement in atopic dermatitis often refers to the flexural surfaces, not typically the extensor surfaces like in psoriasis.
    Less Indicative/Contradictory:
    If classic hallmark features of Psoriasis (e.g., silvery scales, sharply demarcated plaques on extensors), Lichen Planus (e.g., polygonal papules, oral involvement), or PRP (e.g., follicular papules, orange hue) are strongly present, those diagnoses are usually favored over a less specific "chronic dermatitis."
    Age: Atopic dermatitis often begins in childhood but can persist or start in adulthood. Other forms can occur at any age.
    Pityriasis Rubra Pilaris (PRP):
    Key Indicators:
    follicular_papules: Hallmark feature - hyperkeratotic papules centered on hair follicles, giving a "nutmeg grater" texture (high values like 2-3 are very suggestive). Often on dorsal fingers, elbows, knees.
    erythema: Distinctive orange-red or salmon-colored hue. (Interpret erythema value with this color in mind).
    "Islands of sparing": Unaffected skin within larger areas of redness (not a direct input feature but a classic sign).
    scalp_involvement: Common, often with diffuse erythema and yellowish scaling.
    Palmoplantar keratoderma: Thickening of skin on palms and soles (high scaling on extremities might hint at this, but it's not a specific input).
    definite_borders: Involved areas can be sharply demarcated from normal skin.
    knee_and_elbow_involvement: Common sites for follicular_papules and erythema/scaling.
    itching: Variable.
    Less Indicative/Contradictory:
    Absence of follicular_papules (0) is a strong argument against PRP.
    Absence of the characteristic orange-red hue (though erythema only gives intensity).
    Presence of polygonal_papules.
    Age: Bimodal age distribution: first two decades and then in the 50s-60s.
    General Analytical Strategy for Probability Estimation:

    Holistic Review: Do not assess features in isolation. Consider the entire clinical picture.
    Hallmark Features: Give significant weight to the presence (high score) or absence (score of 0) of pathognomonic or highly characteristic features for each disease.
    Consistency: Check for consistency across features. For example, if knee_and_elbow_involvement is high, does it fit the typical extensor pattern of psoriasis or the follicular papules of PRP on these sites?
    Differential Diagnosis: Actively consider why it might be one disease and not another. For instance, if itching is high, it could be lichen planus or chronic dermatitis; polygonal_papules would then strongly steer towards lichen planus. If scaling is high, is it silvery (psoriasis-like), greasy (seborrheic dermatitis-like), or collarette-like (pityriasis rosea-like)? (The input only gives intensity, so make reasonable inferences where the pattern of other features supports a type of scaling).
    Age and Family History: Use age as a modifying factor (e.g., pityriasis rosea is less common in older patients, PRP has bimodal peaks). Use family_history primarily for psoriasis and atopic forms of chronic dermatitis.
    Probabilities: The probabilities should reflect your confidence in each diagnosis based on the evidence. They should sum to 1.0 (or be interpretable as relative likelihoods that can be normalized). A disease that perfectly matches its classic profile with multiple high-scoring key features and no contradictions should receive a high probability. A disease with many absent key features or contradictory findings should receive a very low probability.
    Example of applying the logic (Hypothetical Case - do not use these probabilities for the actual case below):
    Consider: 'erythema: 3, scaling: 3, definite_borders: 3, itching: 1, koebner_phenomenon: 2, polygonal_papules: 0, follicular_papules: 0, oral_mucosal_involvement: 0, knee_and_elbow_involvement: 3, scalp_involvement: 2, family_history: 1, age: 30'

    Psoriasis: Many features align strongly (high erythema, scaling, definite borders, knee/elbow, scalp, Koebner + family history, appropriate age). High probability.
    Seborrheic Dermatitis: While scalp involvement is present, the other features (thick scale, very definite borders, prominent knee/elbow) are less typical for SD as the primary picture. Lower probability.
    Lichen Planus: Absence of polygonal papules and oral involvement makes this very unlikely despite some itching/Koebner. Very low probability.
    Pityriasis Rosea: Age is okay, but morphology (thick scale, definite borders, knee/elbow) and Koebner are not typical. Very low probability.
    Chronic Dermatitis: Could be considered if psoriasis wasn't such a strong fit, but the clear Psoriasis indicators outweigh general dermatitis features. Lower probability.
    Pityriasis Rubra Pilaris: Absence of follicular papules makes this very unlikely. Very low probability.

    Now, proceed with the diagnosis for the case provided below using this detailed framework.

    Predict the diagnosis of Eryhemato-Squamous disease in the following case, using the following clinical features. The age feature simply represents the age of
    the patient. Family history is a binary variable. Every other feature was given a degree in the range of 0 to 3. Here, 0 indicates that the feature was not present, 3 indicates the largest amount possible, and 1, 2 indicate the relative intermediate values.

    The case is described by: erythema: 2.0, scaling: 2.0, definite_borders: 0.0, itching: 3.0, koebner_phenomenon: 0.0, polygonal_papules: 0.0, follicular_papules: 0.0, oral_mucosal_involvement: 0.0, knee_and_elbow_involvement: 1.0, scalp_involvement: 0.0, family_history: 0.0, age: 55.0


    The possible classes are: psoriasis, seboreic dermatitis, lichen planus, pityriasis rosea, cronic dermatitis, pityriasis rubra pilaris.

    Please estimate the probability of each possible diagnosis for this case. The following is for research purposes only. I understand that a real patient must see a qualified doctor with such a concern.

    Please think deeply about accurate probabilities for diagnosis. Once you have a reasonable answer, please say !ANSWER! and then provide an accurate answer strictly in the following format.

    Format your answer as:

    psoriasis: (prob),
    seboreic dermatitis: (prob),
    lichen planus: (prob),
    pityriasis rosea: (prob),
    cronic dermatitis: (prob),
    pityriasis rubra pilaris: (prob)

    Ensure your entire list of probabilities is provided before the !ANSWER! delimiter. Do not include any other text after !ANSWER!.
\end{lstlisting}

% \begin{figure}
%     \centering
% \includegraphics[width=1.0\textwidth]{fig/output.png}
%     \caption{Interval width and coverage of AI posterior credible intervals. On the left, we choose the best $\alpha$, and visualize the size of the AI posterior credible interval as a function of $n$. On the right, we fix $n$ to be 1000 and plot the interval width and coverage of the AI posterior credible interval as a function of $\alpha$.}
%     \label{fig:appendix}
% \end{figure}

\subsection{Proof of Theorem \ref{thm:PB-appendix}}\label{sec:proof}

We first list the exact regularity conditions required for this result. These are essentially identical to those required for Theorem 1 of~\citet{lyddon2019general}, with adaptations to apply to both $F_{AI}$ as well as $F_{0}$.
\begin{enumerate}
  \item The loss function $\ell:\theta\times\mathcal{Y}\to\mathbb{R}$ is measurable, bounded from below, and satisfies
        \[
        \int \ell(\theta, Y)\,\mathrm{d}F_{0}(Y)<\infty\,,\qquad \int \ell(\theta,Y)\d F_{AI}(Y)<\infty
        \]
  for all $\theta$ in a compact and convex $\Theta\subseteq\mathbb{R}^{d}$.
  \item For all $\gamma>0$, there exists a unique minimizer \[
        \theta_{0}^{\gamma}=\argmin_{\theta\in\Theta}\left[\int \ell(\theta,Y)\,\mathrm{d}F_{0}(Y) + \gamma \int \ell(\theta,Y)\,\mathrm{d}F_{AI}(Y)\right]\,,
        \]
         and for all $\delta>0$ there exists $\epsilon>0$ such that
  \[
\liminf_{n} P\left(\sup_{\lvert\theta-\theta_{0}^{\gamma}\rvert\geq\delta} \frac{1}{n}\sum_{i=1}^{n}\left[\ell(\theta,Y_{i})-\ell(\theta_{0}^{\gamma},Y_{i})\right]\geq\epsilon\right)=1\,.
  \]
\item For each $\gamma>0$, there exists an open ball $B$ such that $\theta_{0}^{\gamma}\in B$, in which for almost all $Y\in\mathcal{Y}$ the first three partial derivatives of $\ell(\theta,Y)$ with respect to $\theta\in B$ exist and are continuous. In addition, there exist measurable functions $G_{j}$, $G_{jk}$, $G_{jkl}$, $H_{jkl}$ such that for $\theta\in B$,
\begin{align*}
\left|\frac{\partial\ell(\theta,x)}{\partial\theta_j}\right| &\leq G_j(Y) \quad \text{where} \quad \int G_j(Y) \, dF_0(Y) + \int G_j(Y) \, dF_{AI}(Y)< \infty, \\
\left|\frac{\partial^2\ell(\theta,Y)}{\partial\theta_j\partial\theta_k}\right| &\leq G_{jk}(Y) \quad \text{where} \quad \int G_{jk}(Y) \, dF_0(Y) + \int G_{jk}(Y) \, dF_{AI}(Y) < \infty, \\
\left|\frac{\partial^3\ell(\theta,Y)}{\partial\theta_j\partial\theta_k\partial\theta_l}\right| &\leq G_{jkl}(Y) \quad \text{where} \quad \int G_{jkl}(Y) \, dF_0(Y) + \int G_{jkl}(Y) \, dF_{AI}(Y) < \infty, \\
\left|\frac{\partial\ell(\theta,Y)}{\partial\theta_j}\frac{\partial^2\ell(\theta,Y)}{\partial\theta_k\partial\theta_l}\right| &\leq H_{jkl}(Y) \quad \text{where} \quad \int H_{jkl}(Y) \, dF_0(Y) + \int H_{jkl}(Y) \, dF_{AI}(Y) < \infty.
\end{align*}
                                                                                        \item The matrices
\begin{align*}
I_{1}(\theta) =\int\nabla \ell(\theta,Y)\nabla(\theta,Y)^{\top}\d F_{0}(Y)\,,&\qquad  I_{2}(\theta) =\int\nabla \ell(\theta,Y)\nabla(\theta,Y)^{\top}\d F_{AI}(Y) \\
J_{1}(\theta) =\int\nabla^{2} \ell(\theta,Y)\d F_{0}(Y)\,,&\qquad  J_{2}(\theta) =\int\nabla^{2} \ell(\theta,Y)\d F_{AI}(Y)
\end{align*}
are all positive definite for $\theta\in B$ with all elements finite.
\end{enumerate}

\begin{proof}
We assume that all of the aforementioned regularity conditions hold. Define the weighted generalized score function by
  \[
\wt{S}_{n}^{\alpha}=\sum_{i=1}^{n}w_{i}\nabla\ell(\theta,Y_{i})+\sum_{j=1}^{m} \wt{w}_{j}\nabla\ell(\theta, Y^{*}_{j})\,,
  \]
  and similarly the weighted sample generalized information matrix by
  \[
    \wt{J}_{n}^{\alpha}(\theta) = \sum_{i=1}^{n} w_{i} \nabla^{2} \ell(\theta,Y_{i}) + \sum_{j=1}^{m} \wt{w}_{j} \nabla^{2} \ell(\theta, Y^{*}_{j})\,.
  \]
  where $(w_{1},\ldots,w_{n},\wt{w}_{1},\ldots,\wt{w}_{m})'\sim\mathrm{Dirichlet}(1,\ldots,1,\alpha/m,\ldots,\alpha/m)$. We first argue that $(n+\alpha)^{1/2}\wt{S}_{n}^{\alpha}(\wh{\theta}_{n}^{\alpha})\overset{d}{\to} N(0, I(\theta_{0}^{\gamma}))$ using the Cram\'{e}r-Wold device as follows. Let $z\in \mathbb{R}^{p}$ with $\|z\|_{1}=1$, and define
  \begin{align*}
    t_{n,\alpha}(z) &\equiv \sqrt{n+\alpha}\, z^{\top} \wt{S}_{n}^{\alpha}(\wh{\theta}_{n}^{\alpha}) \\
    &= \sqrt{n+\alpha}\,\sum_{k=1}^{p} z_{k} \left(\frac{\sum_{i=1}^{n} V_{i} g_{k}(\wh{\theta}_{n}^{\alpha},Y_{i})+\sum_{j=1}^{m} \wt{V}_{j}g_{k}(\wh{\theta}_{n}^{\alpha},Y^{*}_{j}) }{\sum_{i=1}^{n}V_{i}+\sum_{j=1}^{m} \wt{V}_{j}}\right)
  \end{align*}
  where $V_{1},\ldots,V_{n}\overset{iid}{\sim}\mathrm{Exp}(1)$, $\wt{V}_{1},\ldots,\wt{V}_{m}\overset{iid}{\sim}\mathrm{Gam}(\alpha/m,1)$, and $g_{k}(\theta',Y)\equiv \frac{\partial \ell(\theta,Y)}{\partial \theta_{k}}\rvert_{\theta=\theta'}$. This can be rewritten as
  \[
t_{n,\alpha}(z)=\frac{1}{(n+\alpha)^{-1}\left(\sum_{i}V_{i}+\sum_{j}\wt{V}_{j}\right)}\cdot \frac{\sum_{i=1}^{n} a_{i}V_{i} + \sum_{j=1}^{m} \wt{a}_{j}\wt{V}_{j}}{\sqrt{n+\alpha}}
  \]
  where $a_{i}=\sum_{k=1}^{K} z_{k} g_{k}(\wh{\theta}_{n}^{\alpha},Y_{i})$ and $\wt{a}_{j}=\sum_{k=1}^{K} z_{k}g_{k}(\wh{\theta}_{n}^{\alpha}, Y_{j}^{*})$.
  The first factor in the product converges almost surely to $1$. Thus, it is sufficient to show that
  \[
    t_{n,\alpha}'(z)=\frac{\sum_{i=1}^{n} a_{i}V_{i} + \sum_{j=1}^{m} \wt{a}_{j}\wt{V}_{j}}{\sqrt{n+\alpha}}
  \]
  converges in distribution to $N(0,z^{\top} I(\theta_{0}^{\gamma})z)$. Indeed, we appeal to the Lindeberg-Feller-L\'{e}vy CLT. The variance of the expression is given by
  \[
\bar{\sigma}^{2}_{n,\alpha} = (n+\alpha)^{-1}\left(\sum_{i=1}^{n} a_{i}^{2} + \frac{\alpha}{m}\sum_{j=1}^{m} \wt{a}_{j}^{2}\right)
  \]
  and the asymptotic covariance is thus
  \[
    \lim_{n\to\infty} \bar{\sigma}^{2}_{n,\alpha}= z^{\top}\left(\frac{I_{1}(\theta_{0}^{\gamma})+\gamma I_{2}(\theta_{0}^{\gamma})}{1+\gamma}\right)z = z^{\top}I(\theta_{0}^{\gamma})z\,.
  \]
  Therefore, we have $\sqrt{n+\alpha}\, \wt{S}_{n}^{\alpha}(\wh{\theta}_{n}^{\alpha})\overset{d}{\to} N(0,I(\theta_{0}^{\gamma})$.

  Assuming smoothness conditions \citep{lyddon2019general, Newton1991}  hold, we can perform a Taylor expansion of the weighted score function around the empirical risk minimizer $\wh{\theta}_{n}^{\alpha}$ as
  \[
    \wt{S}_{n}^{\alpha}(\wh{\theta}_{n}^{\alpha}) = (\wt{J}_{n}(\wh{\theta}_{n}^{\alpha}) - R_{n})(\theta^{*}-\wh{\theta}_{n}^{\alpha})
  \]
  for a remainder term $R_{n}$. Following~\citet{lyddon2019general}, $\wt{J}_{n}^{\alpha}(\wh{\theta}_{n}^{\alpha})-R_{n}$ converges to $J(\theta_{0}^{\gamma})$ and is invertible with high probability. Slutsky's theorem then yields the desired result, with asymptotic covariance given by $J(\theta_{0}^{\gamma})^{-1}I(\theta_{0}^{\gamma})J(\theta_{0}^{\gamma})^{-1}$.

  {
  \paragraph{Continuous base measure.} When the base measure $F_{AI}$ is continuous, the posterior bootstrap algorithm approximates it using $m$ i.i.d. samples $Z_1, \ldots, Z_m \sim F_{AI}$. We consider the asymptotic regime where $m/n\to r$ for some constant $r>0$, and $\alpha = \gamma n$. We denote the weighted generalized score function and sample generalized information matrix as $\wt{S}_{n,m}^{\alpha}$ and $\wt{J}_{n,m}^{\alpha}$ respectively, defined analogously to the atomic case:
  \[
\wt{S}_{n,m}^{\alpha}(\theta)=\sum_{i=1}^{n}w_{i}\nabla\ell(\theta,Y_{i})+\sum_{j=1}^{m} \wt{w}_{j}\nabla\ell(\theta, Z_{j})\,,
  \]
  \[
    \wt{J}_{n,m}^{\alpha}(\theta) = \sum_{i=1}^{n} w_{i} \nabla^{2} \ell(\theta,Y_{i}) + \sum_{j=1}^{m} \wt{w}_{j} \nabla^{2} \ell(\theta, Z_{j})\,,
  \]
  where $(w_{1},\ldots,w_{n},\wt{w}_{1},\ldots,\wt{w}_{m})'\sim\mathrm{Dirichlet}(1,\ldots,1,\alpha/m,\ldots,\alpha/m)$. Let $\wh\theta_{n,m}^\alpha$ be the empirical risk minimizer obtained using $Y_1, \ldots, Y_n$ and $Z_1, \ldots, Z_m$. Under standard regularity conditions for M-estimation, $\wh\theta_{n,m}^\alpha \xrightarrow{p} \theta_0^\gamma$ as $n,m\to\infty$, where $\theta_0^\gamma$ remains is defined as in~\eqref{eq:oracle-riskmin}.

  We use the Cramér-Wold device as in the proof of Theorem~\ref{thm:PB-appendix}. We define $t_{n,m,\alpha}(z)$ analogously using $\wt{S}_{n,m}^{\alpha}(\wh\theta_{n,m}^\alpha)$ where $V_i \sim \text{Exp}(1)$ and $\wt{V}_j \sim \text{Gam}(\alpha/m, 1)$. Since $\alpha/m \approx \gamma n / (rn) = \gamma/r$, the first parameter of the latter Gamma distribution converges to $\gamma/r$. As established previously, the factor $(\sum V_k + \sum \wt{V}_l)/(n+\alpha)$ converges almost surely to one. It remains to argue that
  \[
    t'_{n,m,\alpha}(z)=\frac{\sum_{i=1}^{n}a_{i}V_{i}+\sum_{j=1}^{m}\wt{a}_{j}\wt{V}_{j}}{\sqrt{n+\alpha}}
  \]
  converges in distribution as $n\to\infty$ to $N(0, z^{\top}I(\theta_{0}^{\gamma})z)$, where $a_i = z^\top\nabla\ell(\wh\theta_{n,m}^\alpha, Y_i)$ and $\wt{a}_j = z^\top\nabla\ell(\wh\theta_{n,m}^\alpha, Z_j)$.

  The variance of $t'_{n,m,\alpha}(z)$, conditional on the data $\{Y_i, Z_j\}$, is
  \[
\bar{\sigma}^{2}_{n,m,\alpha} = \frac{1}{n+\alpha}\left(\sum_{i=1}^{n} a_{i}^{2}\text{Var}(V_i) + \sum_{j=1}^{m} \wt{a}_{j}^{2}\text{Var}(\wt{V}_j)\right) = \frac{n}{n+\alpha}\left(\frac{1}{n}\sum_{i=1}^{n} a_{i}^{2}\right) + \frac{m(\alpha/m)}{n+\alpha}\left(\frac{1}{m}\sum_{j=1}^{m} \wt{a}_{j}^{2}\right)\,.
  \]
  As $n,m \to \infty$, since $\wh\theta_{n,m}^\alpha \xrightarrow{p} \theta_0^\gamma$, under regularity conditions we have
  \begin{align*} \frac{1}{n}\sum_{i=1}^{n} a_{i}^{2} &= \frac{1}{n}\sum_{i=1}^{n} (z^\top\nabla\ell(\wh\theta_{n,m}^\alpha, Y_i))^2 \xrightarrow{p}  z^\top I_1(\theta_0^\gamma) z \\ \frac{1}{m}\sum_{j=1}^{m} \wt{a}_{j}^{2} &= \frac{1}{m}\sum_{j=1}^{m} (z^\top\nabla\ell(\wh\theta_{n,m}^\alpha, Z_j))^2 \xrightarrow{p} z^\top I_2(\theta_0^\gamma) z \end{align*}
  Noting that $\text{Var}(\wt{V}_j) \to \gamma/r$ and using the limits $\frac{n}{n+\alpha} \to \frac{1}{1+\gamma}$ and $\frac{m(\alpha/m)}{n+\alpha} = \frac{\alpha}{n+\alpha} \to \frac{\gamma}{1+\gamma}$, the asymptotic variance is
  \[
  \lim_{n\to\infty} \bar{\sigma}^{2}_{n,m,\alpha} = \frac{1}{1+\gamma} (z^\top I_1(\theta_0^\gamma) z) + \frac{\gamma}{1+\gamma} (z^\top I_2(\theta_0^\gamma) z) = z^\top I(\theta_0^\gamma) z\,,
  \]
  where $I(\theta)$ is defined as $\frac{I_1(\theta)+\gamma I_2(\theta)}{1+\gamma}$.
  As in~\citet{lyddon2019general}, the Lindeberg condition for the sum (conditional on the data) holds under regularity conditions on $\ell$, ensuring conditional convergence via the Lindeberg-Feller CLT. Since the limiting variance $z^\top I(\theta_0^\gamma) z$ does not depend on the specific data sequence, this implies the unconditional convergence: $t'_{n,m,\alpha}(z) \overset{d}{\to} N(0, z^\top I(\theta_0^\gamma) z)$. The Cramér-Wold device then yields, $\sqrt{n+\alpha}\, \wt{S}_{n,m}^{\alpha}(\wh{\theta}_{n,m}^{\alpha})\overset{d}{\to} N(0,I(\theta_{0}^{\gamma}))$.

  Assuming smoothness conditions \citep{lyddon2019general, Newton1991}  hold, we can perform a Taylor expansion of the weighted score function around the empirical risk minimizer $\wh{\theta}_{n,m}^{\alpha}$ as
  \[
    \wt{S}_{n,m}^{\alpha}(\wh{\theta}_{n,m}^{\alpha}) = (\wt{J}_{n,m}(\wh{\theta}_{n,m}^{\alpha}) - R_{n,m})(\theta^{*}-\wh{\theta}_{n,m}^{\alpha})
  \]
  for a remainder term $R_{n,m}$. Following~\citet{lyddon2019general}, $\wt{J}_{n,m}^{\alpha}(\wh{\theta}_{n,m}^{\alpha})-R_{n,m}$ converges in probability to $J(\theta_{0}^{\gamma})$ and is invertible with high probability. Slutsky's theorem then yields the desired result once again, with asymptotic covariance given by $J(\theta_{0}^{\gamma})^{-1}I(\theta_{0}^{\gamma})J(\theta_{0}^{\gamma})^{-1}$.}
\end{proof}

\end{document}